\documentclass{aa}
\usepackage{epsfig}
\usepackage{amsmath}
\usepackage{txfonts}
\usepackage{natbib}
\bibpunct{(}{)}{;}{a}{}{,} 
\usepackage{bm}

\usepackage{subcaption}
\usepackage{colortbl}\usepackage[colorlinks,linkcolor=blue,citecolor=blue,linktocpage=true,breaklinks, plainpages=false,urlcolor=blue]{hyperref}
\usepackage{graphicx}
\mathchardef\mhyphen="2D
\usepackage[tracking=true,kerning=true,shrink=50]{microtype}
\usepackage{gensymb}
\usepackage{textcomp}

\def\Halpha{\mbox{H\hspace{0.1ex}$\alpha$}}

\def\FeI{\ion{Fe}{i}}
\def\CaII{\ion{Ca}{ii}}

\begin{document}

\title{Coronal kink oscillations and photospheric driving}
\subtitle{combining SolO/EUI and SST/CRISP high-resolution observations}

\author{N. Poirier\inst{1,2}, S. Danilovic \inst{3}, P. Kohutova\inst{1,2}, C. J. Díaz Baso \inst{1,2}, L. Rouppe van der Voort \inst{1,2}, D. Calchetti \inst{4}, J. Sinjan \inst{4}}
\institute{Rosseland Centre for Solar Physics, University of Oslo, P.O. Box 1029, Blindern, NO-0315 Oslo, Norway\\
\email{nicolas.poirier@astro.uio.no}
\and
Institute of Theoretical Astrophysics, University of Oslo, P.O. Box 1029, Blindern, NO-0315 Oslo, Norway
\and
Institute for Solar Physics, Dept. of Astronomy, Stockholm University, Albanova University Center, 10691 Stockholm, Sweden
\and
Max-Planck-Institut für Sonnensystemforschung, Justus-von-Liebig-Weg 3, 37077 Göttingen, Germany}
\date{Received 15 November 2024; accepted 22 February 2025}

\abstract
{The driving and excitation mechanisms of decay-less kink oscillations in coronal loops remain under active debate. The photospheric dynamics may provide the continuous energy supply required to sustain these oscillations.}
{We aim to quantify and provide simple observational constraints on the photospheric driving of coronal loops in a few typical active region configurations: sunspot, plage, pores and enhanced-network regions. We then aim to investigate the possible interplay between the photospheric driving and the properties of kink oscillations in the connected coronal loops.}
{We analyse two unique datasets of the corona and photosphere taken at a high spatial and temporal resolution during the first coordinated observation campaign between Solar Orbiter and the Swedish 1-m Solar Telescope (SST). A local correlation tracking method is applied on the SST/CRISP data to quantify the photospheric motions at the base of coronal loops. The same loops are then analysed in the corona by exploiting data from the Extreme Ultraviolet Imager (EUI) on Solar Orbiter, and by using a wavelet analysis to characterize the detected kink oscillations.}
{Each type of photospheric region shows varying dynamics with an overall increase in strength going from pore, plage, enhanced-network to sunspot regions. Differences can also be seen in the amplitudes of the fundamental kink mode measured in the corresponding coronal loops. This suggests the photosphere is involved in the driving of coronal kink oscillations. However, the few samples available does not allow to further establish the excitation mechanism yet.}
{Despite oscillating coronal loops being anchored in seemingly "static" strong magnetic field regions as seen from coronal EUV observations, photospheric observations provide evidence for a continuous and significant driving at their base. The precise connection between photospheric driving and coronal kink oscillations remains to be further investigated. Upcoming coordinated observations between Solar Orbiter and ground-based telescopes will provide crucial additional observational constraints, with this pilot study serving as a baseline for future works. This study finally provides critical constraints on both the quasi-steady and broadband photospheric driving that can be tested in existing numerical models of coronal loops.}


\titlerunning{Coronal kink oscillations and photospheric driving}
\authorrunning{Poirier et al.}

\maketitle


\section{Introduction}
\label{sec:intro}
Transverse oscillations in coronal loops have long been observed by near Earth observatories and more recently with Solar Orbiter \citep{Muller2020}. Although their properties are quite well-known now, their driver and excitation mechanism remain under active debate \citep[see e.g. the review by ][]{Nakariakov2021}.

First observed by \citet{Nakariakov1999,Aschwanden1999a} with the \emph{Transition Region And Coronal Explorer} \citep[TRACE, ][]{Handy1999}, a wealth of transverse oscillations has been detected in active region loops. They were then routinely observed with the \emph{Atmospheric Imaging Assembly} \citep[AIA, ][]{Lemen2012} on board the \emph{Solar Dynamics Observatory} \citep[SDO, ][]{Pesnell2012} with the majority of transverse oscillations being interpreted as standing kink waves \citep{Anfinogentov2015}. More generally, the solar corona was found to be dominantly filled by transverse oscillatory power as seen in Doppler velocity maps obtained with the \emph{Coronal Multi-channel Polarimeter} \citep[CoMP, ][]{Tomczyk2008}.

Two main regimes of transverse kink oscillations in coronal loops were identified. Short-lived kink oscillations have often been detected due to their large amplitude, although they usually last for a few oscillation periods only with an envelope that decays (super-)exponentially \citep[e.g.][]{Nakariakov1999,Aschwanden1999a,White2012,Nistico2013,Goddard2016,Nechaeva2019}. On the other hand, small-amplitude kink oscillations without apparent decay (decay-less) became later routinely observed with the advent of SDO/AIA \citep{Wang2012,Tian2012,Nistico2013,Anfinogentov2013,Anfinogentov2015,Zhong2022a} and more recently with the \emph{Extreme Ultraviolet Imager} \citep[EUI, ][]{Rochus2020} on board Solar Orbiter \citep[e.g.][]{Mandal2022,Li2023,Petrova2023,Zhong2023,Shrivastav2024}. 

The large-amplitude kink oscillations have often been associated with impulsive coronal events such as reconnection events during solar flares \citep[e.g.][]{Aschwanden1999a,Nakariakov1999,White2012,Nistico2013}, but the driver for the small-amplitude decay-less regime remains unknown. That driver must provide a continuous input of energy to sustain the coronal oscillations for at least a few oscillation cycles. Catastrophic condensation events known as coronal rain have been found to trigger transverse kink oscillations in coronal loops \citep{Verwichte2017,Kohutova2017,Shrivastav2024b}. Coronal rain observed periodically \citep[the coronal monsoon, see ][]{Auchere2018} may be a promising quasi-continuous driver of coronal origin. The photosphere is another potential driver (if not the most obvious) with its never ending convective motions and substantial energy reservoir. In this paper we investigate the potential of photospheric driving to excite and sustain kink oscillations in coronal loops. \\

The paper is structured as follows. We first give some context and background in section \ref{sec:background}. We then introduce the observations used and the data analysis methods in section \ref{sec:data-method}. We then present the results of the photospheric and coronal analyses in section \ref{sec:results_photosphere} and \ref{sec:results_corona} respectively, which we then combine together and discuss in section \ref{sec:results_photosphere-corona}. Limitations of the methodology as well as perspectives for future work are discussed in section \ref{sec:discussion}. We finally summarise the key conclusions in section \ref{sec:conclusions}.

\section{Background}
\label{sec:background}
\subsection{Potential sources and types of photospheric drivers}
\label{sec:intro_drivers}

Photospheric drivers of different types and origins have been investigated over the last decade, either quasi-harmonic, random (broadband) or quasi-steady in nature.

Quasi-harmonic drivers have shown potential to trigger kink oscillations \citep{Ballai2008,Selwa2010,Nistico2013,Karampelas2017,Pagano2017,Afanasyev2019,Riedl2019,Guo2019,Guo2023,Gao2023b}, which can be justified by the observed ubiquitous leakage of the global photospheric five-minute p-modes into the transition region \citep{Gao2023a} and corona \citep[see e.g.][]{Ballai2008,Morton2016,Morton2019}. The p-modes (for pressure-modes) originate within the interior of the sun and more specifically in the convection zone which acts as a resonant cavity \citep{Ulrich1970}. However, harmonic drivers are found to be the most effective when their frequency is close enough to the natural kink-mode frequencies of the oscillating coronal loops \citep{Ballai2008,Selwa2010}, and the driver frequency (and its harmonics) also tend to be dominant over the kink mode \citep{Ballai2008}. Statistical studies of kink oscillations show overall a clear linear relationship between the observed period and length of the oscillating loops \citep{Anfinogentov2013,Nistico2013,Anfinogentov2015,Mandal2022,Zhong2022a,Li2023}. A harmonic driving from the photospheric p-modes may only be efficient in a subset of loops with compatible resonant frequencies as we will discuss later, and so cannot explain the whole spectrum of observed oscillating coronal loops.

Measurements of velocity fluctuations in the corona by CoMP also suggest that a significant part of the oscillating power must be generated by stochastic or random processes \citep{Morton2016}. This is highly indicative of the important role of photospheric convective motions, from granular to super-granular scales, in the generation of coronal oscillations. Such scenario have been tested in simulations by applying random (broadband) drivers to simulated coronal loops \citep{Pagano2019a,Afanasyev2020,Ruderman2021,Karampelas2024}. Most of them manage to simulate kink oscillations with striking similarities with actual observations; however, they fail at producing the observed linear polarisation of kink modes \citep{Zhong2023}. Furthermore, a variation of the slope in the oscillating power spectrum in the corona has been detected in different magnetic regions observed by CoMP \citep{Morton2016}. This has also been noted in photospheric observations where the convection was seen to be suppressed in locations with strong magnetic field such as plage regions \citep[e.g.][]{Title1989}. The magnetic field is known to have an influence on the velocity power spectrum in the photosphere as seen also in simulations \citep{YellesChaouche2014}. 

Last but not least, constant or quasi-steady drivers have also been investigated in analogy with the vibration of a violin string in response to a moving bow \citep[aka the bow-on-a-string model, ][]{Goedbloed1995,Nakariakov2016a,Karampelas2020}. Unlike the other two drivers mentioned above, quasi-steady drivers have the advantage to agree with the observed dominant linear polarisation of kink modes \citep{Zhong2023}. However, the simulated kink oscillations take a long time to develop and hence the excitation requires long lasting flows which could be hard to justify from an observational perspective. Additionally, the simulated decay-less kink oscillations tend to have too low amplitudes compared to observations \citep{Karampelas2020}. In the case of coronal loops anchored in active sunspots, such flows could be associated with the strong moat flows that continuously propagate outwards from the penumbrae \citep{Lohner-bottcher2013,Strecker2018}. More generally, magnetic elements in the photosphere are observed to be systematically transported at meso/super-granular scales \citep[e.g.][]{Orozcosuarez2012,Malherbe2017}. Systematic flows or motions at large scales, whether associated with super-granular flows or sunspots, have been extensively detected in photospheric observations and discussed in the literature \citep[see the review by ][and references therein]{Rincon2018}. They are known to operate on timescales of hours at least or even over the whole lifetime of active regions \citep{Strecker2018}, which is much longer than the time required for the kink oscillations to establish in the simulations by \citet{Karampelas2020}. \\

The bow-on-a-string excitation mechanism of coronal loop kink oscillations is sometimes mistakenly thought to require steady flows to operate. The important point is that the model applies to any low-frequency driving motion that occurs at timescales longer than the kink-mode period, and that covers a broad spectrum of the solar granulation and magnetic flux transport dynamics. Furthermore, this excitation mechanism has been shown to work well in the presence of noise \citep{Nakariakov2022}, where the latter can be considered as an additional broadband random component to the driving as we will discuss throughout this paper. Another debate comes from the fact that the footpoints of coronal loops often appear as steady in (E)UV images due to their strong anchor in photospheric regions with high magnetic flux concentration. However the background coronal magnetic field surrounding the coronal loops is likely not as steady and gets systematically dragged along with the aforementioned photospheric motions and transport of magnetic flux at all scales. In summary the relative, and not absolute, motions are a key element in the bow-on-a-string interpretation of the excitation of coronal kink oscillations.

\subsection{Excitation mechanism: forced or self oscillator?}
\label{sec:background_excitation}

The photospheric driving of coronal kink oscillations can be seen as a two-fold process. First, the driving force that applies on the tied points of the loop in the photosphere, and second the feedback interaction that occurs at higher heights and that results from the loop moving through the background plasma and magnetic field. 

If only the first component was to be considered, the loop system can be qualified as a forced oscillator as defined in \citet{Jenkins2013}. In that case, the photospheric driving would be the most efficient if it is quasi-harmonic with a frequency that is close enough to one of the resonant frequencies of the loop (fundamental and harmonics). This condition may be satisfied in coronal loops with a compatible length, due to the significant overlap in frequency between the five-minute photospheric p-modes and the observed frequency of kink oscillations \cite[see e.g.][fig 7]{Aschwanden1999a}. Furthermore, the coupling of photospheric p-modes with coronal kink oscillations has been suggested in both observations \citep{Gao2023a} and simulations \citep{Gao2023b}. However, the forced-oscillatory scenario may not explain the whole spectrum of oscillating coronal loops given the large diversity of their observed properties in terms of shape, density and anchor region.

This is where the second component of the photospheric driving, the feedback interaction of the loop with the background corona, becomes essential. Once the loop footpoints are put in motion the surrounding plasma will oppose some resistance in the higher parts of the loop. At some point magnetic tension will operate to straighten up the loop and put it back to "equilibrium". This essentially results in a stick-slip or \citet{Helmholtz1954}'s interaction where the loop alternatively "sticks" with the relatively moving background and "slips" back due to the restoring force \citep[magnetic tension, ][]{Nakariakov2016a}. This produces a force that is periodically aligned with the kink oscillations and hence feed them. Such systems are qualified as self oscillators because they sustain themselves as long as the driver velocity is fast enough and is steady on timescales longer than the resonant period of the system \citep{Jenkins2013}. Self oscillators have the key advantage that they oscillate at their own natural resonant frequency independent of the driver. In other words, self oscillators can turn non-periodic and low-frequency driving into resonant oscillations. Coronal loops are therefore likely to behave as self oscillators for any photospheric driving that occurs at time scales longer than the kink-mode oscillations, such as the convective motions and transport of magnetic flux that occur at scales larger than granules. The self oscillator amplitude grows exponentially compared to the forced oscillator, which grows linearly, however they also rapidly reach a saturation limit when non-linear effects start to develop naturally \citep{Jenkins2013}. Simulations with continuous footpoint driving also show such saturation \citep{Karampelas2019} and hence also agree with observations. For instance vortices typically form at the boundary of simulated oscillating loops as a result of the Kelvin-Helmholtz instability (KHi) \citep[see e.g.][and references therein]{Antolin2016,Antolin2014}, which produce out-of-phase flows that damp and saturate the amplitudes of the oscillations. Resonant absorption is also known to occur in these boundary layers with a transfer of the kink-mode energy into the local Alfvén waves \citep{Pascoe2010,Pascoe2012,Pascoe2013}, from where the energy eventually ends up being thermally dissipated to the plasma due to turbulence and phase-mixing \citep[e.g.][]{Antolin2016}. \\

Both the forced- and self-oscillatory excitation behaviours are expected to manifest in oscillating coronal loops. If the photospheric driver operates at timescales comparable or longer than that of the kink-mode, then either the forced- or self-oscillatory behaviour will dominate respectively. Furthermore, if decay-less kink oscillations are excited by a broadband random spectrum related to the photospheric convection, then a variation of the oscillation properties is expected for coronal loops that connect to different magnetic regions such as sunspots, pores, plages or enhanced networks. These two questions constitute the primary motivations for this work.


\section{Data and method}
\label{sec:data-method}

\subsection{Context: a first coordinated campaign between Solar Orbiter and the Swedish 1-m Solar Telescope}
\label{sec:data-method_2023campaign}

\begin{figure*}
    \centering
    \includegraphics[width=2.\columnwidth]{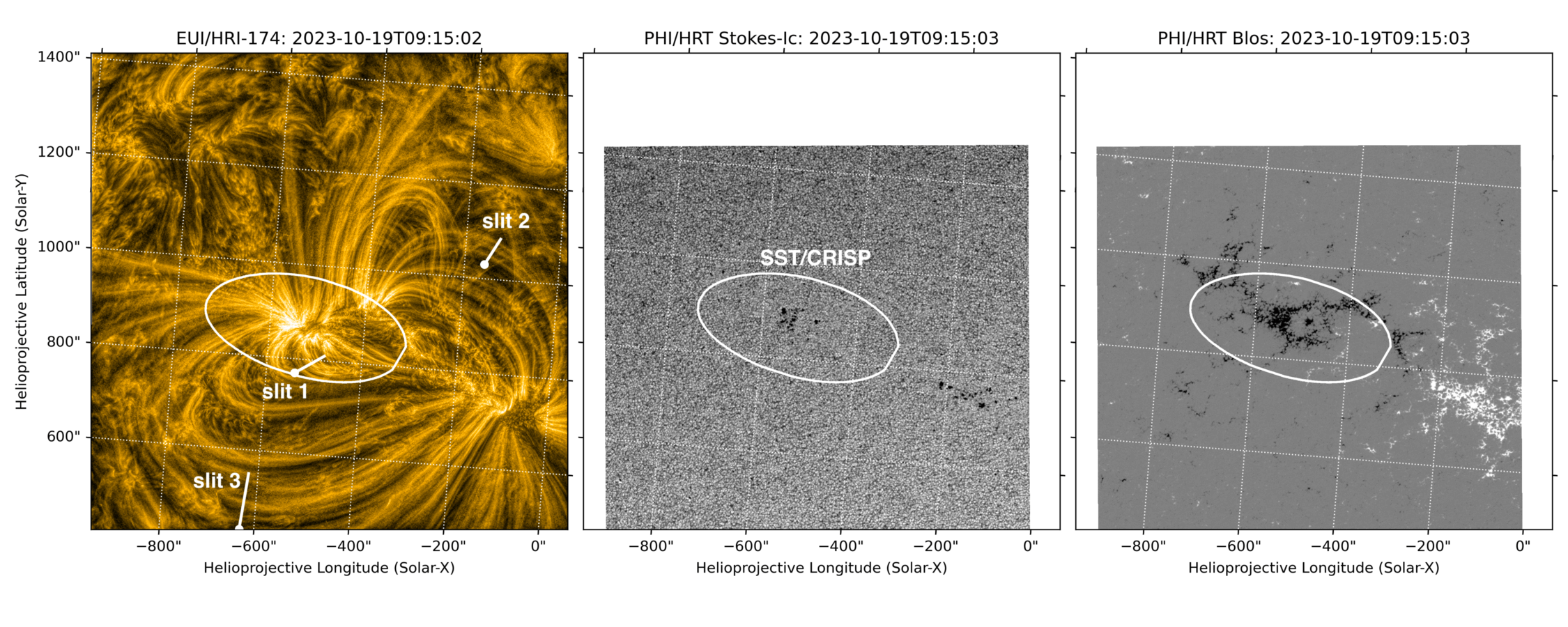}
    \includegraphics[width=2.\columnwidth]{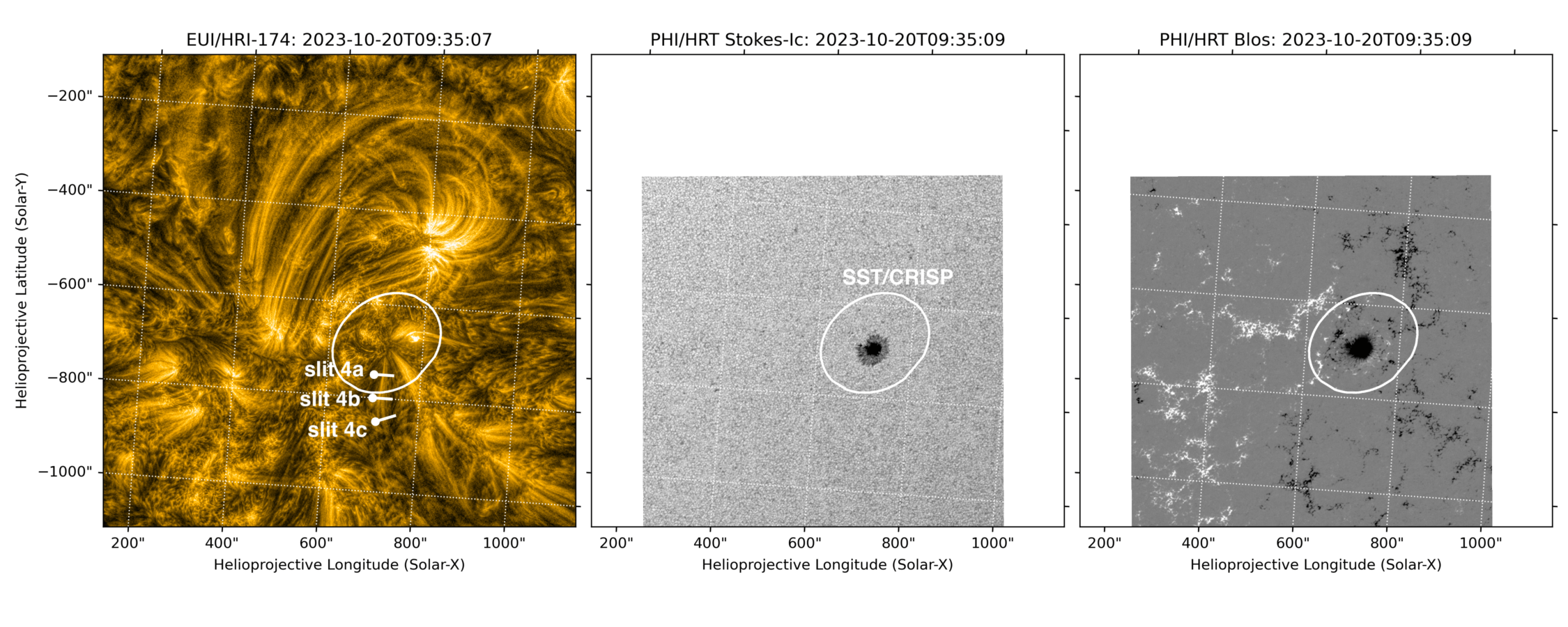}
	\caption{Context view of AR13470 (top row) and AR13468 (bottom row) observed on 19 and 20 Oct 2023 respectively. Left column: WoW-enhanced EUI/HRI-174\AA images with the slits (white lines) used for the coronal oscillation analysis. Middle column: Stokes-I continuum from PHI/HRT. Right column: PHI/HRT inverted line-of-sight component of the magnetic field, linearly scaled between -500\,G (black colours) to 500\,G (white colours). The white ellipse shows the re-projected SST/CRISP FOV. The thin white dashed lines represent the helioprojective coordinate frame as seen from Solar Orbiter.}
	\label{fig:obs_context}
\end{figure*}

In order to investigate the photosphere-corona connection, we exploit two unique datasets acquired during the first coordinated campaign between Solar Orbiter \citep[SolO;][]{Muller2020} and the Swedish 1-m Solar Telescope \citep[SST;][]{Scharmer2003} on October 2023 \citep[see also][]{Danilovic2024}. 
The separation angle between SolO and Earth was 42-45\textdegree for October 19-20, which allowed to pick targets on the Sun visible from both SolO and Earth. For the first time, coronal imaging from the \emph{High Resolution Imager} of EUI \citep[EUI/HRI, ][]{Rochus2020} on Solar Orbiter can be combined with the high-resolution observing capabilities of the SST that provides full spectro-polarimetric diagnostics from the photosphere to the chromosphere. Two distinct active regions were targeted.

Active region NOAA 13470 (hereafter called AR13470) observed on 19 Oct 2023 had a complex morphology where multiple loop systems can be seen as shown in Fig.~\ref{fig:obs_context} (top panel). We focus particularly on the northern part of this active region which was observed by SST/CRISP. The High Resolution Telescope (HRT) of the Polarimeter and Helioseismic Imager (PHI) on Solar Orbiter \citep{Gandorfer2018,Solanki2020} indicates that the coronal loops connect mostly to solar plage and pore regions in the photosphere.

Active region NOAA 13468 (hereafter called AR13468) observed on 20 Oct 2023 includes a sunspot in its core as shown in Fig.~\ref{fig:obs_context} (bottom panel). A large loop arcade is visible in the EUI/HRI image north of the sunspot. Unfortunately, the footpoints of these loops are located just outside of the SST/CRISP field-of-view (FOV). There are nonetheless a few loop footpoints that connect to the sunspot, for which the coronal oscillation properties are analysed in section \ref{sec:results_corona_sunspot}. The photospheric dynamics around the sunspot is analysed with SST/CRISP in section \ref{sec:results_photosphere_sunspot}.

\begin{table*}[t]
	\caption{Summary of the observation datasets used}
	\label{tab:data}
	\centering
	\tabcolsep=0.11cm
	\begin{tabular}{ccccccc}
		\hline\hline
		Active Region & Date/Time & Source & Channel & Type & Cadence & Spatial resolution\\ 
		  (NOAA)  & [UT] & & [\AA] & & [sec] & [km] \\
		  \hline
		13470  & 19 Oct 2023 09:05-10:06 & SST/CRISP & $6173\ (\rm{\ion{Fe}{i}})$ & SP & $27$ & $\approx 110$ (diffraction-limited)\\
                             & 19 Oct 2023 09:00-10:08 & EUI/HRI & $174\ (\rm{\ion{Fe}{ix}})$ & I & $6$ & $\approx 300$ (2-pixel) \\
		13468 & 20 Oct 2023 09:05-10:51 & SST/CRISP & $6563\ (\rm{H_\alpha)}$ & S & $\approx 9$ & $\approx 150$ (diffraction-limited)\\
                             & 20 Oct 2023 09:00-10:59 & EUI/HRI & $174\ (\rm{\ion{Fe}{ix}})$ & I & $10$ & $\approx 300$ (2-pixel)\\
		\hline
	\end{tabular}
	\tablefoot{I, S and SP stand for imaging, spectroscopy and spectropolarimetry respectively.}
\end{table*}

\subsection{Photosphere}
\label{sec:data-method_photosphere}

We exploit high-spatial resolution observations of the photosphere taken by the \emph{CRisp Imaging SpectroPolarimeter} \citep[CRISP;][]{Scharmer2008} at the SST \citep[see also][for the latest upgrades]{Scharmer2019}. CRISP provides high-quality spectral diagnostics of the photosphere and chromosphere in the \FeI~6173~\AA, \Halpha~6563~\AA, and \CaII~8542~\AA\ spectral lines with high-sensitivity polarimetry allowing for accurate magnetic field inference. Together with the other instruments (CHROMIS, \citealt{Scharmer2008}; and MiHI, \citealt{vanNoort2022a}), this makes the SST a highly-versatile ground-based observatory that is highly relevant for the analysis of many solar features at high precision, and has proven to be very effective in coordination with space-based observations \citep[see, e.g.,][]{2014Sci...346D.315D,Antolin2015,Froment2020,Rouppe2020}.

While its one metre aperture is not the largest in solar telescopes, the SST provides observations of unprecedented high quality thanks to exceptional observatory site characteristics, a cutting-edge adaptive optics system \citep{Scharmer2024}, multiple instrument upgrades brought over about twenty years of operations \citep[see][]{Scharmer2019}, and a well-established data-reduction pipeline \citep[SSTRED;][]{DeLaCruz2015,Lofdahl2021} which includes image restoration using the multi-object multi-frame blind deconvolution method \citep[MOMFBD;][]{vanNoort2005}. All together, this allows CRISP to detect small-scale features close to the diffraction limit of $\approx 110\mhyphen 150\ \rm{km}$. Depending on the selected spectral lines and sampling program, the cadence for CRISP typically varies from $\approx 10$ to $\approx 40\ \rm{s}$ allowing the observation of highly dynamical features as well. When full spectro-polarimetric observations of \FeI~6173~\AA\ were taken with CRISP, we inferred the full vector of the photospheric magnetic field using a Milne-Eddington inversion code from \citet{DeLaCruz2019}. While having a smaller FOV than EUI/HRI, CRISP had a recent upgrade of its FOV up to $\approx$87\arcsec\ allowing to study larger-scale dynamics.

In this paper we will use CRISP time series whose specifications are summarised in Table~\ref{tab:data}. Apparent motions in the photosphere (and within the plane of the image) are tracked within the wide-band images of \FeI~6173~\AA (or \Halpha~6563~\AA) using a local correlation tracking \citep[LCT, ][]{November1988} method. The data preparation includes precise co-alignment, removal of blurry frames if necessary, and p-modes filtering. A subset of motion patterns (trajectories) corresponding to the footpoint regions of coronal loops are then selected using a simple threshold on the line-of-sight (LOS) component of the magnetic field.

Lastly, we quantify the extracted photospheric motions with simple parameters that can be directly used as constraints for coronal loop simulations. Along each traced trajectory (for either the full FOV or a sub selection), we first compute the temporal average of the horizontal velocity over the whole time series $\Bar{v_h}$ which corresponds to the constant or quasi-steady component of the photospheric driving. We then compute the temporal 1-D Fourier spectrum of the complex horizontal velocity defined as $v_x+j*v_y$, and we fit two power laws to the power spectral density (PSD) (using a least-square algorithm available in the \emph{lmfit} \emph{Python} library). We extract the fitted slopes $a_1$ and $a_2$ for the low and high frequency components of the Fourier spectrum respectively, as well as the cut-off frequency $f_c$. All four photospheric parameters are summarised in table \ref{tab:results_photosphere}.

\subsection{Corona}
\label{sec:data-method_corona}

We use high spatial and temporal resolution coronal imaging observations in the $174\ \AA$ EUV channel from the \emph{High Resolution Imager} of EUI \citep[EUI/HRI, ][]{Rochus2020} on Solar Orbiter, taken during a first coordinated campaign with the SST in October 2023. EUI/HRI observations consist of a dataset taken on 19 Oct 2023 with 68~min duration and 6~s cadence; and a dataset taken on 20 Oct 2023 with 119~min duration and 10~s cadence (see Table \ref{tab:data}). EUI/HRI has a plate scale of 0\farcs49~pixel$^{-1}$ that can resolve spatial structures as small as $\approx 200\ \rm{km}$ at closest distance ($0.28\ \rm{au}$). Taking a heliocentric distance of $0.4\ \rm{au}$ for the two datasets used in this study, the spatial resolution slightly increases up to $\approx 300\ \rm{km}$. The EUI/HRI FOV is about $1000\arcsec$ wide and hence covers large areas such as active regions. We exploit level-2 data from the data release 6.0 \citep{euidatarelease6}. 

All data preparatory steps including most of the data analyses have been done within the \emph{Sunpy} open-source \emph{Python} ecosystem \citep{Sunpy2020}. The preparation step includes co-alignment and small-scale feature enhancing using the wavelet-optimized whitening \citep[WoW;][]{Auchere2023} method. Artificial slits are placed across the coronal loops and near their apex when possible, along which the EUI/HRI intensities are extracted and projected into a time-distance map format. The intensities are averaged over a few pixels across the slits axis to improve the signal-to-noise ratio. Oscillating loops are then tracked over time using a multi-Gaussian fitting method. The fitted loop centres are then de-trended with a high-pass Fourier filter to retrieve the displacement amplitude of the kink-mode oscillations. A wavelet transform is then applied to get the dominant periods within the oscillatory signals. The results of the wavelet transform are finally reduced in clusters within which the final coronal oscillation parameters are extracted, that is the average displacement amplitude and period associated with the detected kink-mode oscillations. For more details about the method please refer to Appendix \ref{sec:appendix_method_corona}.

\section{Results: photospheric motions}
\label{sec:results_photosphere}
We start our analysis investigating the connectivity of coronal loops to different photospheric regions and their particular properties.

\subsection{AR13470: pores, plages and enhanced-network}
\label{sec:results_photosphere_Oct19}

\begin{figure*}
    \centering
	\includegraphics[width=1.72\columnwidth]{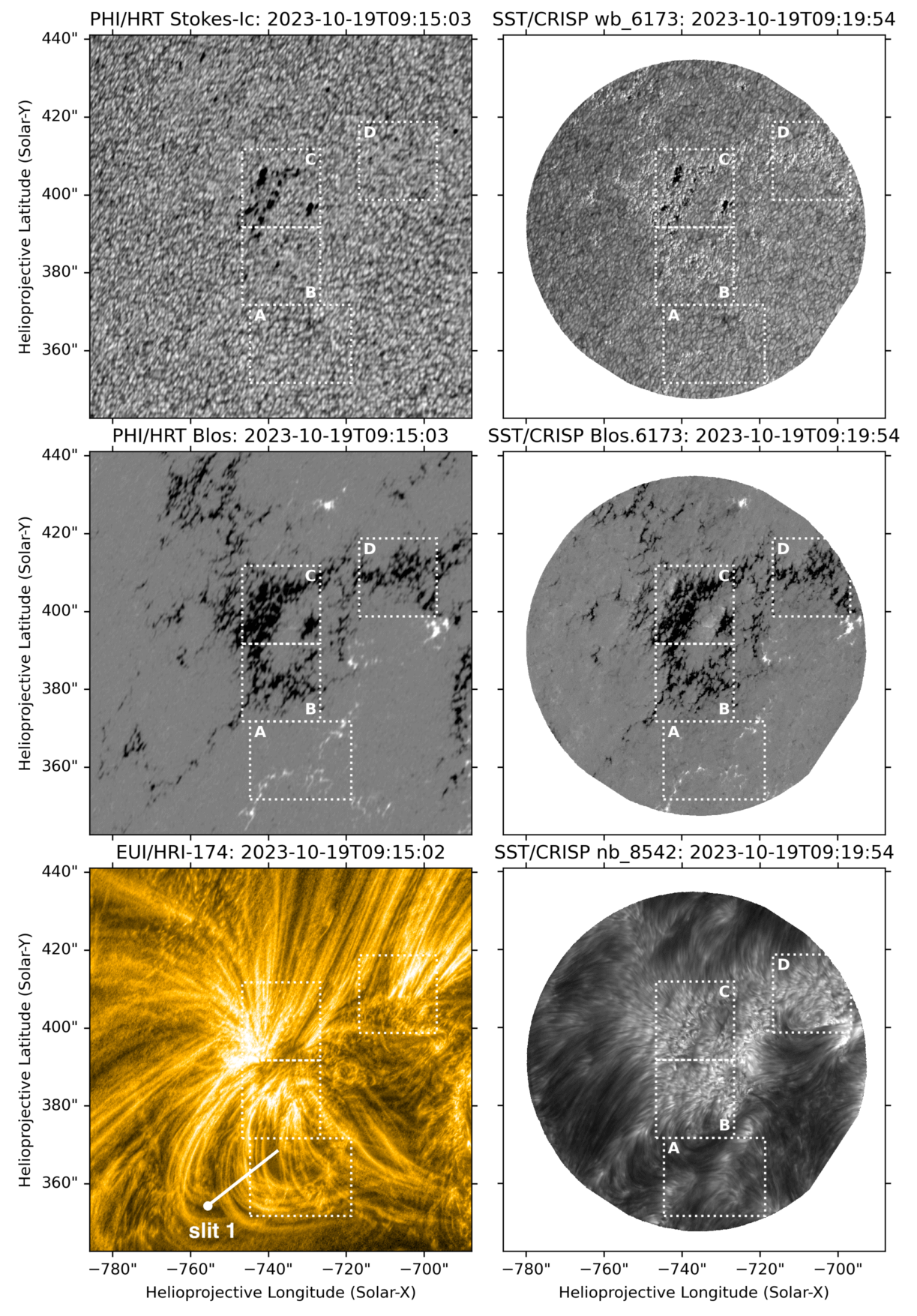}
    \caption{Cutout centred on the core of AR13470. Left column: PHI/HRT and EUI/HRI observations re-projected on the SST/CRISP frame. Right column: SST/CRISP observations for the \FeI~6173~\AA continuum (wide-band filter, top), inverted LOS magnetic field (middle) and line core of \CaII~8542~\AA (bottom). The white rectangles depict the sub-regions selected for the photospheric motion analyses (see text). The continuum intensities are colour-plotted on a logarithmic scale (only the values between the $0.1\%$ ad $99.9\%$ percentiles are mapped). The line-of-sight magnetic field maps are shown on a linear scale ranging -500\,G (black colours) to 500\,G (white colours). The artificial slit traced in white colour is used later for the coronal oscillation analysis.}
	\label{fig:Oct19_context}
\end{figure*}

We focus here on the northern part of AR13470 captured by SST/CRISP. A zoomed contextual view is given in Fig.~\ref{fig:Oct19_context} where the EUI/HRI and PHI/HRT observations have been re-projected on the SST/CRISP frame, taking into account the 42\textdegree\ difference in viewing angle. We divided the SST/CRISP FOV in four sub-regions labelled from A to D. Region C corresponds to the very core of the active region where multiple pores can be seen. Regions B and D are regular active region plages that also contain very strong magnetic fields without the presence of pores. The plages can be seen as bright patches around the active region core in the wide-band SST/CRISP image of \FeI~6173~\AA (upper right panel). Finally, region A is qualified as an enhanced network due to its peripheral location and its weaker magnetic field, and can be seen in the LOS magnetic field maps (middle panels). A bunch of coronal loops connect to each of these four photospheric regions. In the following, we estimate and quantify the strength of the photospheric motions that may affect the excitation of coronal kink oscillations.

\begin{figure}
	\centering
	\includegraphics[width=1.\columnwidth]{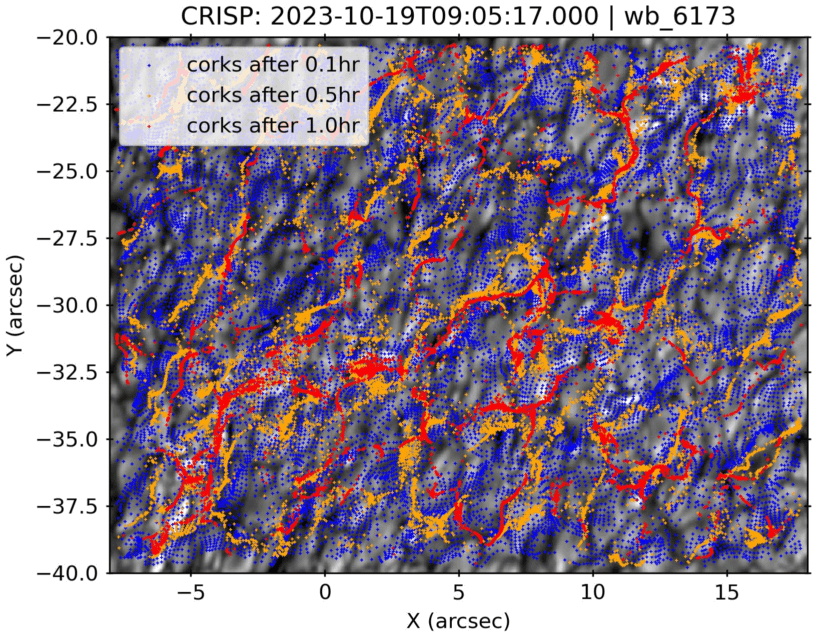}
	\caption{Overview of the LCT-derived motions for region A of AR13470 in the form of corks at different times, for the optimal set of LCT parameters $fwhm=600\ \rm{km}$, $dt=54\ \rm{s}$, over the wide-band image at 6173\,\AA.}
	\label{fig:2023-10-19_corksA}
\end{figure}

The photospheric motion analysis is run within each region following the method described in section \ref{sec:data-method_photosphere} (see also Appendix \ref{sec:appendix_method_photosphere}). By propagating the LCT-derived horizontal velocities over time, one can get an overview of the motions history over the entire time series. An example is shown for region A in Fig.~\ref{fig:2023-10-19_corksA} in the form of corks plotted at different times during the propagation. The corks, which are initially uniformly distributed, get progressively organised into a large-scale network with time. First, the magnetic field rapidly accumulates on small scales within the inter-granular lanes. A slower migration then operates by transporting the magnetic flux further out on super-granular scales until a "barrier" is reached. This barrier can be either the edge of super granules, the enhanced network or the plages where the magnetic flux starts being significant. Such transport on large (super-granular) scales can be a key element in the excitation of kink oscillations in the connected coronal loops, as we will discuss later.

\begin{figure}
	\centering
	\includegraphics[width=1.\columnwidth]{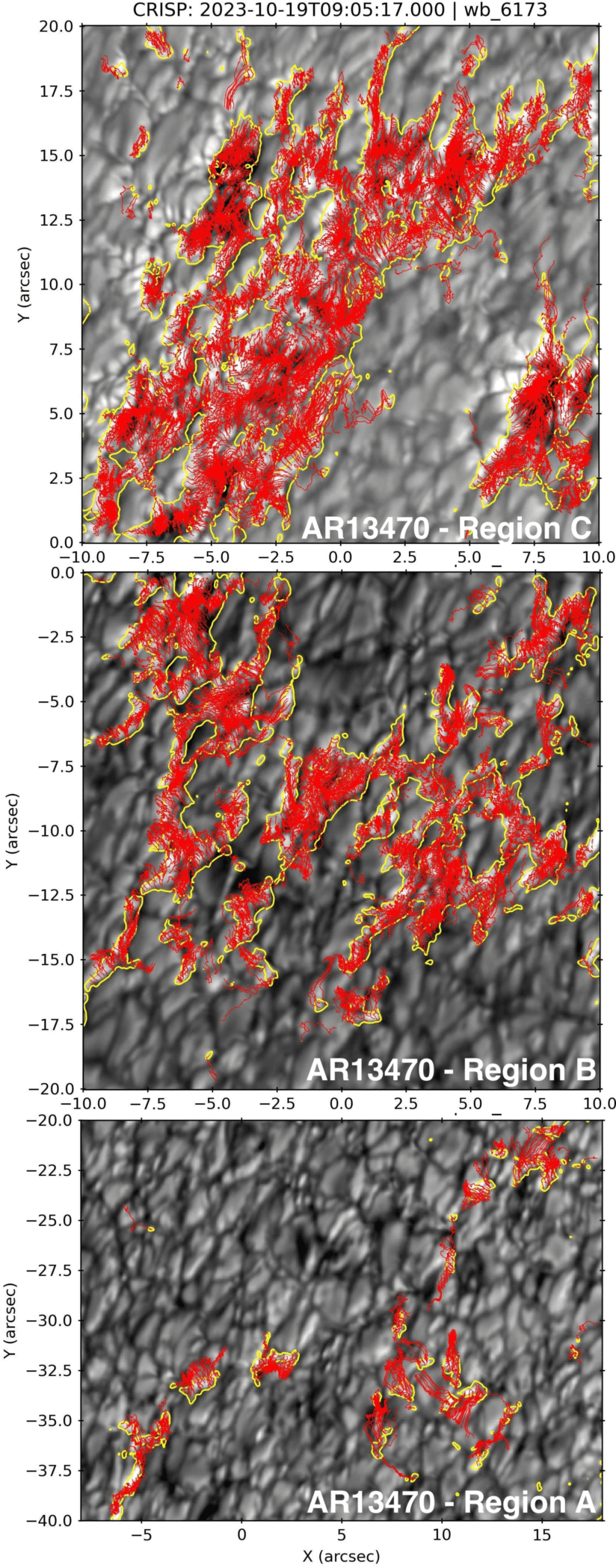}
	\caption{The LCT-derived selected trajectories for region A (bottom), B (middle) and C (top) of AR13470, for the optimal set of LCT parameters $fwhm=600\ \rm{km}$, $dt=54\ \rm{s}$, over the wide-band image at 6173\,\AA.}
	\label{fig:2023-10-19_tracksABC}
\end{figure}

As we want to quantify the photospheric driving at the footpoint of connecting coronal loops, we select among all LCT-derived trajectories the ones associated with the enhanced network (region A) and the plages (region B, C and D) that we identify using a threshold of $B_{\rm LOS}>100\ \rm{G}$ and $B_{\rm LOS}<-200\ \rm{G}$ respectively. The subsets of selected trajectories are shown in Fig.~\ref{fig:2023-10-19_tracksABC} for illustration purposes. The plage in region D is not shown because of its similarity with region B (all results can be found in Table~\ref{tab:results_photosphere}). An additional step is required to further reduce all these trajectories into simple parameters.

\begin{figure*}
	\centering
	\includegraphics[width=0.975\columnwidth]{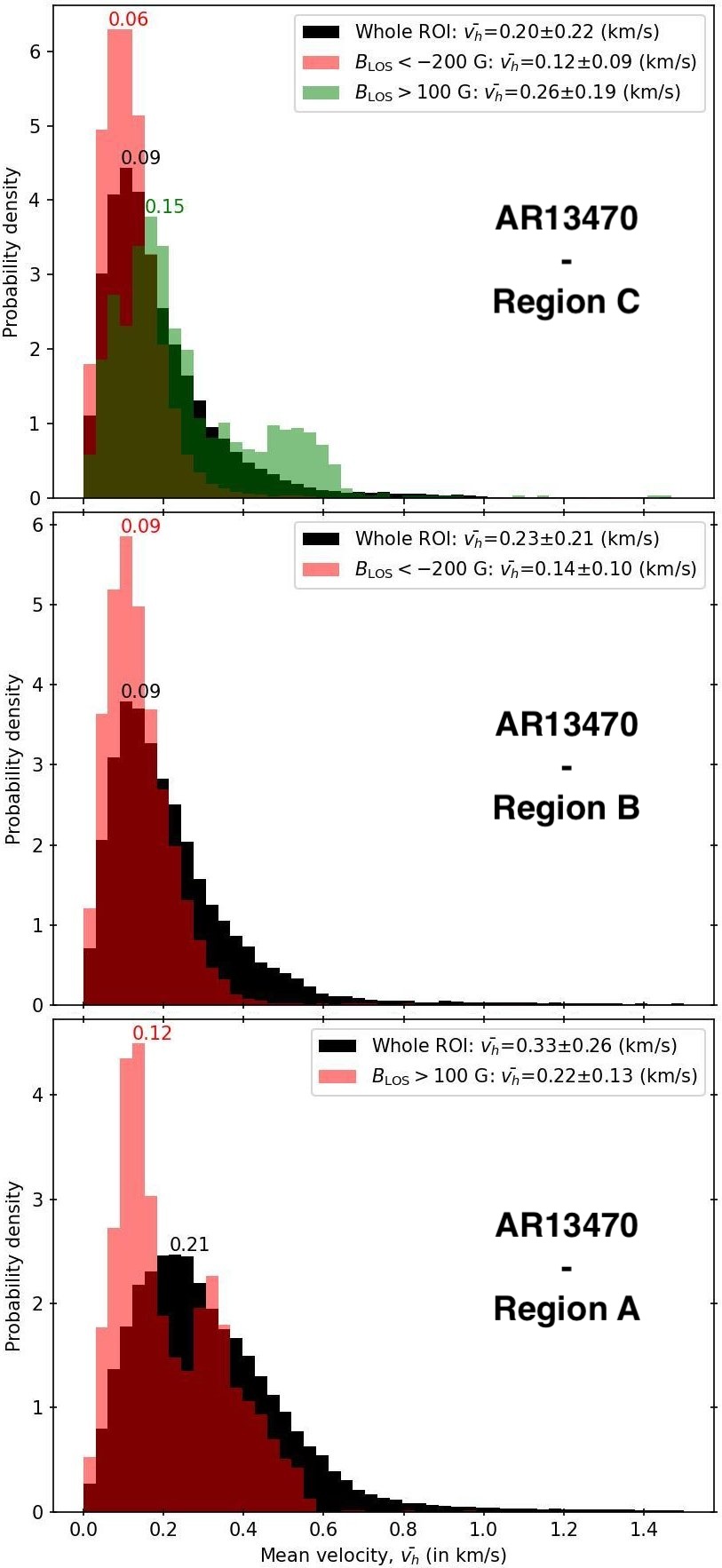}
    \includegraphics[width=1.\columnwidth]{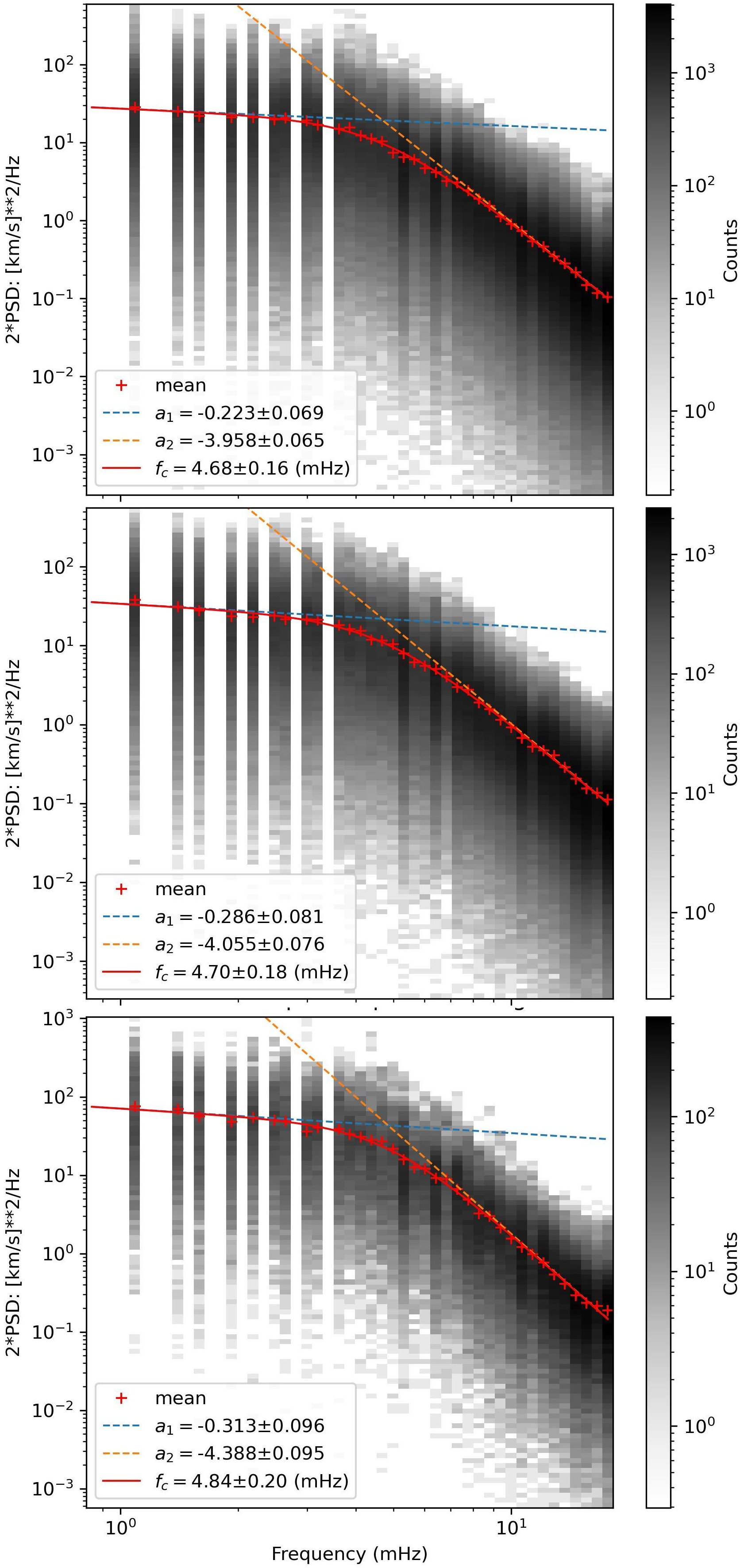}
	\caption{Photospheric motion analysis on AR13470 for the SST/CRISP sub-region A (bottom row), B (middle row) and C (top row). Left column: distribution of the temporal averages $\Bar{v_h}$ over the whole time series. Right column: Fourier spectra of the power spectral density (PSD) for the sub-selection of trajectories corresponding to the red distributions, colour-plotted as 2-D histograms on a log-scale, and with the two power-law fitting (dashed lines).}
	\label{fig:2023-10-19_PHO_results}
\end{figure*}

The photospheric analyses for region A, B and C of AR13470 are plotted in Fig.~\ref{fig:2023-10-19_PHO_results} for the optimal set of LCT parameters $fwhm=600\ \rm{km}$, $dt=54\ \rm{s}$. The left panel shows the distribution of the LCT-derived horizontal velocities $\Bar{v_h}$ after being propagated and temporally averaged along each trajectory. The black colours show the results for all trajectories within each region whereas the red colours correspond to the subset of selected trajectories associated with stronger magnetic field. For all three regions, a first observation is that all the red-colour distributions are shifted towards lower $\Bar{v_h}$ values down to $\approx$0.05-0.1\, km/s. This is reminiscent of the effect of convection suppression in regions with high magnetic flux concentration \citep[see e.g.][]{Title1989}. Nonetheless there is still a non-negligible population among the selected trajectories of the enhanced network (region A) that has stronger motions at around $0.35\ \rm{km/s}$. Such value is consistent with the migration of magnetic elements at super-granular scales \citep{Orozcosuarez2012}. We also look at motions associated with flux emergence in the core of AR13470 in region C (see Fig.~\ref{fig:Oct19_context}). The emerging flux is tracked by filtering magnetic elements of the opposite polarity with $B_{\rm LOS}>100\ \rm{G}$. Albeit being "parasite" such magnetic elements have still a strong magnetic field that might allow them to interact with neighbouring connecting coronal loops. The actual nature of such interaction and its likelihood are left for future investigation. The flux emergence manifests clearly in the distribution of $v_h$ by an additional population of motions at around $0.5\ \rm{km/s}$ (see the green distribution in Fig.~\ref{fig:2023-10-19_PHO_results}, top panel).

To quantify the properties of high and low frequency motions we calculated the 1D Fourier spectra for each region (see right column of Fig.~\ref{fig:2023-10-19_PHO_results}). All the studied cases can be fitted with two power laws with slopes $a_1$ and $a_2$ for the low (blue dashed line) and high (orange dashed line) frequency part respectively (see Appendix \ref{sec:appendix_method_photosphere} for more details). Such two-fold power spectrum is in agreement with the exhaustive study of solar granulation made by \citet{Malherbe2017}. The high-frequency part in all regions A, B and C suggest a red-noise spectrum that depicts a Brownian-like motion of the solar granulation. On the other hand, a slope $|a_1|<1$ is indicative that the low-frequency motions are dominated by advection. The differences in the Fourier spectra are subtle between each region with just a slight variation of the fitted slopes. The slope at high frequencies $a_2$ progressively flattens out going from region A to C that is going from "weak" to "strong" magnetic fields. This can be interpreted as a decrease in the energy contained at the scale of granules. In a similar manner the low-frequency slope $a_1$ gets also reduced in magnitude from region A to C, suggesting that the advection at large scales beyond the granules at meso/super-granular scales get weaker. This variation of the slopes $a_1$ and $a_2$ is a signature of the partial suppression of convective motions and magnetic field transport near regions with higher magnetic field concentration, that seems to affect primarily the motions at and above the granular scale. In that picture, the very core of AR13470 (region C) with its multiple pores represents the most restrictive case in terms of potential photospheric driving of the adjacent coronal loops. The results for the plage in region D are reported in Table~\ref{tab:results_photosphere} and do not show significant difference overall with the plage in region B, except a slightly increased power in the lower frequencies (i.e., steeper $a_1$). The plage in region D is further away from the active region core and hence may be less influenced by the strong magnetic field around the active region core (region C). Therefore this could leave more freedom for the migration and drift of magnetic elements on long timescales.

\subsection{AR13468: sunspot with moat flow}
\label{sec:results_photosphere_sunspot}

Coronal loops anchored in sunspot areas are often presumed to be the least susceptible to be influenced by photospheric driving, because of their apparent static footpoint as seen in EUV. In this section, we will show that sunspots are highly dynamic environments affecting the adjacent coronal loops, especially in the case of a fully developed and active sunspot such as the one studied in this section.

\begin{figure*}
    \centering
	\includegraphics[width=1.73\columnwidth]{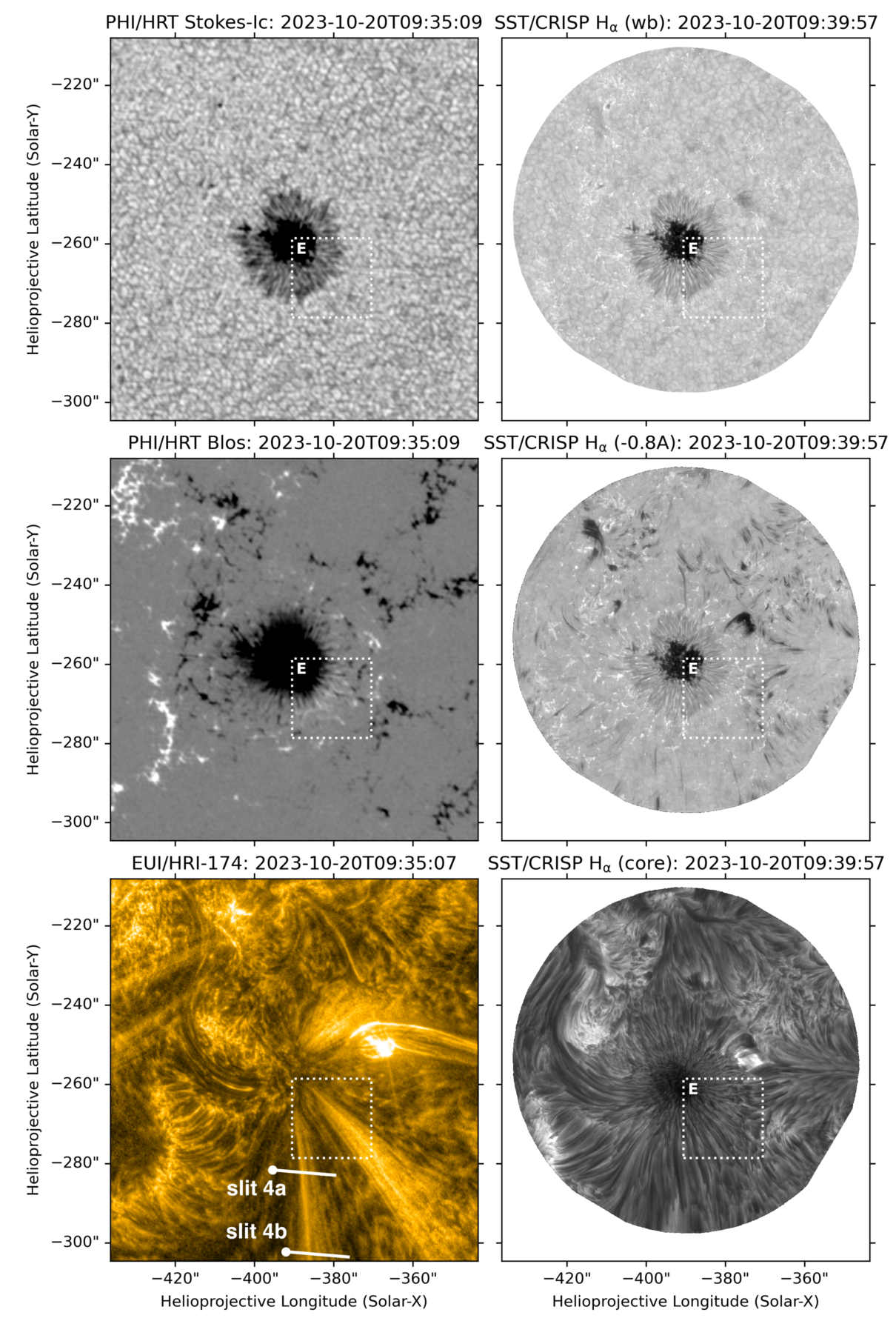}
    \caption{Cutout on the sunspot of AR13468. Left column: PHI/HRT and EUI/HRI observations re-projected on the SST/CRISP frame. Right column: SST/CRISP observations for the \Halpha~6563~\AA\ continuum (wide-band filter, top), $-0.8\AA$ blue wing (middle) and line core (bottom). The white rectangle depicts the sub-region selected for the photospheric motion analysis. The intensities are colour-plotted on a logarithmic scale (only the values between the $0.1\%$ ad $99.9\%$ percentiles are mapped). The artificial slits traced in white colour are used later for the coronal oscillation analysis.}
	\label{fig:Oct20_context}
\end{figure*}

\begin{figure}
    \centering
    \includegraphics[width=.85\columnwidth]{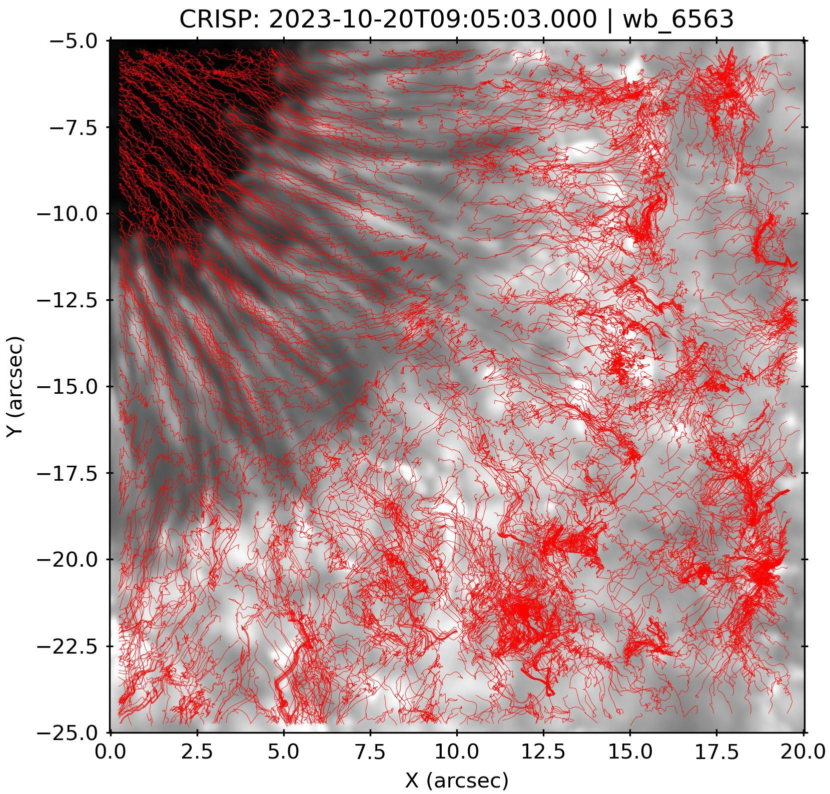}
	\includegraphics[width=.8\columnwidth]{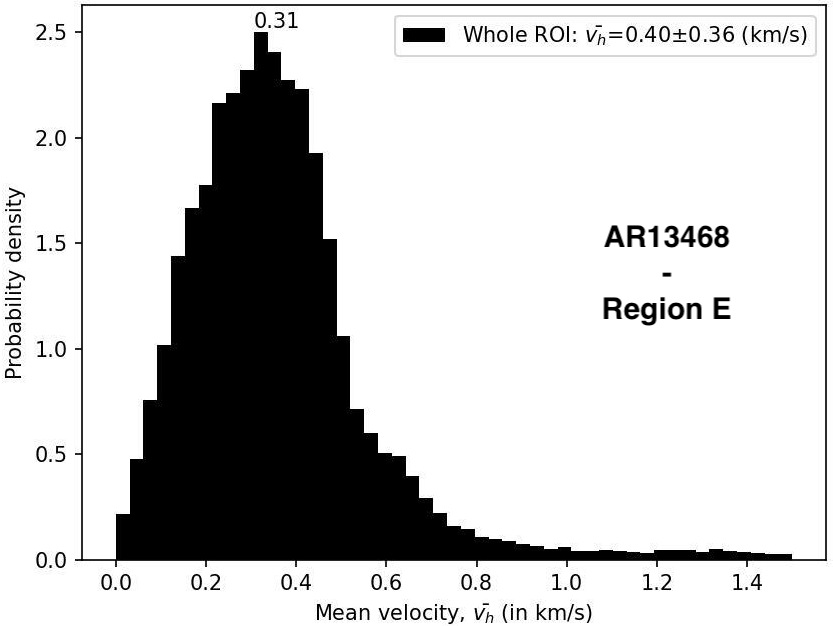}
    \includegraphics[width=.85\columnwidth]{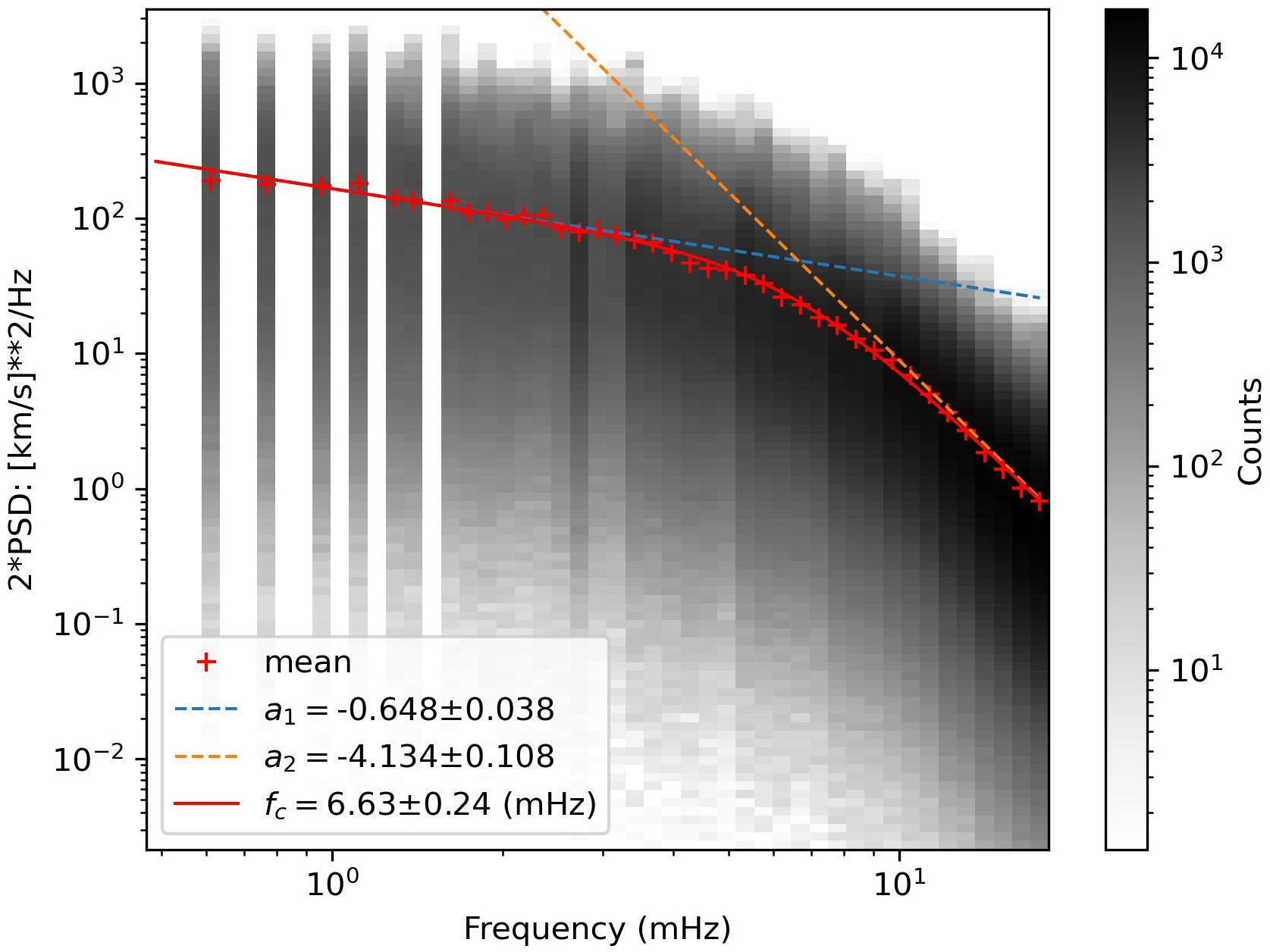}
	\caption{Photospheric motion analysis for the sunspot of AR13468 (region E) where coronal loops are seen to connect. Top panel: trajectories of corks from time $t=0$ and over the full dataset duration (1 hour and 46 min). The corresponding $\Bar{v_h}$-distribution and Fourier spectra along each trajectory are shown in the middle and bottom panels respectively in the same format as in Fig.~\ref{fig:2023-10-19_PHO_results}.}
	\label{fig:2023-10-20_v_results}
\end{figure}

We analyse photospheric motions in and around the sunspot of AR13468 where coronal loops in EUI/HRI are seen to connect. More precisely they connect to its penumbra in region E as illustrated in Fig.~\ref{fig:Oct20_context}. A systematic migration away from the sunspot can be seen in the LCT-derived propagated trajectories as illustrated in the top panel of Fig.~\ref{fig:2023-10-20_v_results}. These motions correspond to the so-called moat flows that are known to surround active sunspots \citep{Lohner-bottcher2013,Strecker2018}. The trajectories start by being mostly radial within the sunspot penumbra. Once injected into the granular network, the motions become Brownian-like due to the small-scale dynamics related to the granulation. On large scales, the motions remain mostly dominated by a constant advection radially away from the sunspot.

Figure~\ref{fig:2023-10-20_v_results} shows the distribution of the average velocity $v_h$ (middle panel) and Fourier spectra (bottom panel) for the LCT-derived motions of region E, using the same format and same optimal set of LCT parameters as in Fig.~\ref{fig:2023-10-19_PHO_results}. The temporal average of the horizontal flows $\Bar{v_h}$ has a distribution that peaks at $0.31\ \rm{km/s}$, the largest in all the datasets examined in this work and is significantly above the quasi-steady velocity driver of $0.3\ \rm{km/s}$ tested in the simulations of \citet{Karampelas2020}. This average was computed over the full duration of the dataset (1 hour and 46 min), but it is known that such flows around sunspots can last even over several days albeit with some little decay \citep{Strecker2018}. This can be further seen in the low-frequency component of the motions with an increased slope $a_1\approx-0.65$ (in magnitude) compared to all the other cases. All of this supports the fact that sunspots, and especially active sunspot with flux emergence or moat flows as shown here, are among the most favourable environments in terms of potential photospheric driving of coronal kink oscillations. The exact nature of this interaction remains to be further investigated, although there are already promising works and prospects in that direction as we will discuss in section \ref{sec:discussion}.

\section{Results: coronal loop oscillations}
\label{sec:results_corona}
We now investigate the oscillating properties of different types of coronal loops depending on their connectivity to the photosphere. If such coronal oscillations are driven by the photosphere, a difference in the properties of these oscillations is expected depending on this connectivity.

\subsection{AR13470: plage/enhanced-network loops}
\label{sec:results_corona_plage-enhanced}

We focus in this section on a bundle of short coronal loops ($L\lesssim 70\ \rm{Mm}$) observed in the core of AR13470 by EUI/HRI. A cutout of the region is shown in Fig.~\ref{fig:Oct19_context}. These loops are very dynamic, show fine structure and connect the plage region B and enhanced-network region A. Here we analyse the oscillations in these loops by making a perpendicular cut close through their apex (slit 1).

\begin{figure*}
    \centering
    \includegraphics[width=1.6\columnwidth]{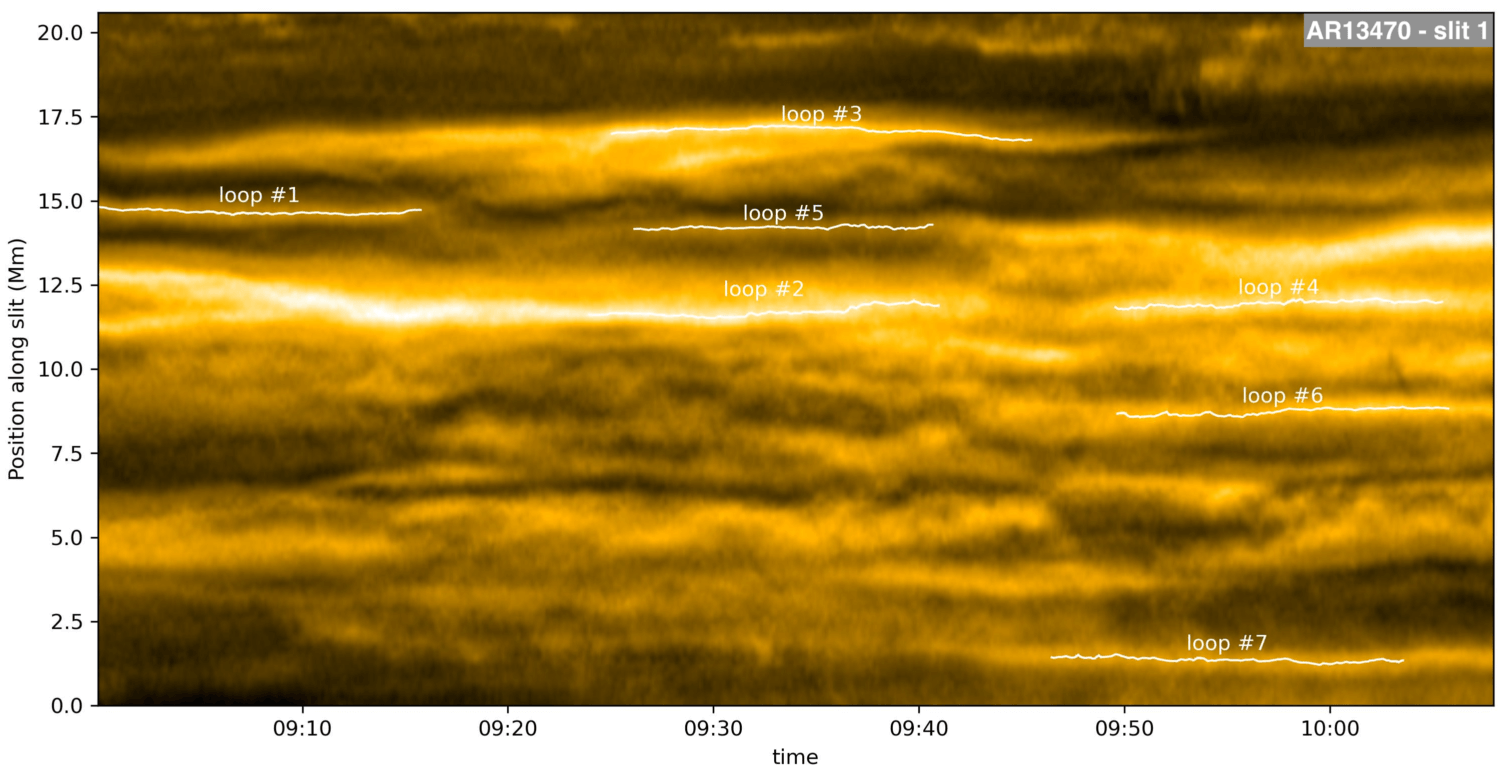}
	\caption{Time-distance map of the EUI/HRI intensity along slit 1 for AR13470, along with the fitted loop centres (white lines). The y-axis represents the distance along the slit axis, starting from the edge marked by a white circle in the EUI/HRI image of Fig.~\ref{fig:Oct19_context}.}
	\label{fig:2023-10-19_plage-enhanced_Imaps}
\end{figure*}

The time-distance intensity map for slit 1 is shown in Fig.~\ref{fig:2023-10-19_plage-enhanced_Imaps}. While a multitude of oscillatory signatures can be seen, the identification of clean oscillation patterns is difficult. Since the kink-mode is global, one would expect some collective behaviour for loops belonging to the same bundle or group. That property is well identified in multi-strand simulations \citep{Luna2019} as well as in observations \citep{Nakariakov1999, Aschwanden1999a, White2012a}. No collective behaviour can clearly be seen in Fig.~\ref{fig:2023-10-19_plage-enhanced_Imaps}. The apparent closeness of these short loops could be the result of a line-of-sight integration effect. A collective behaviour will be more visible in some of the time-distance maps shown later.

\begin{figure*}
    \centering
	\includegraphics[width=1.75\columnwidth]{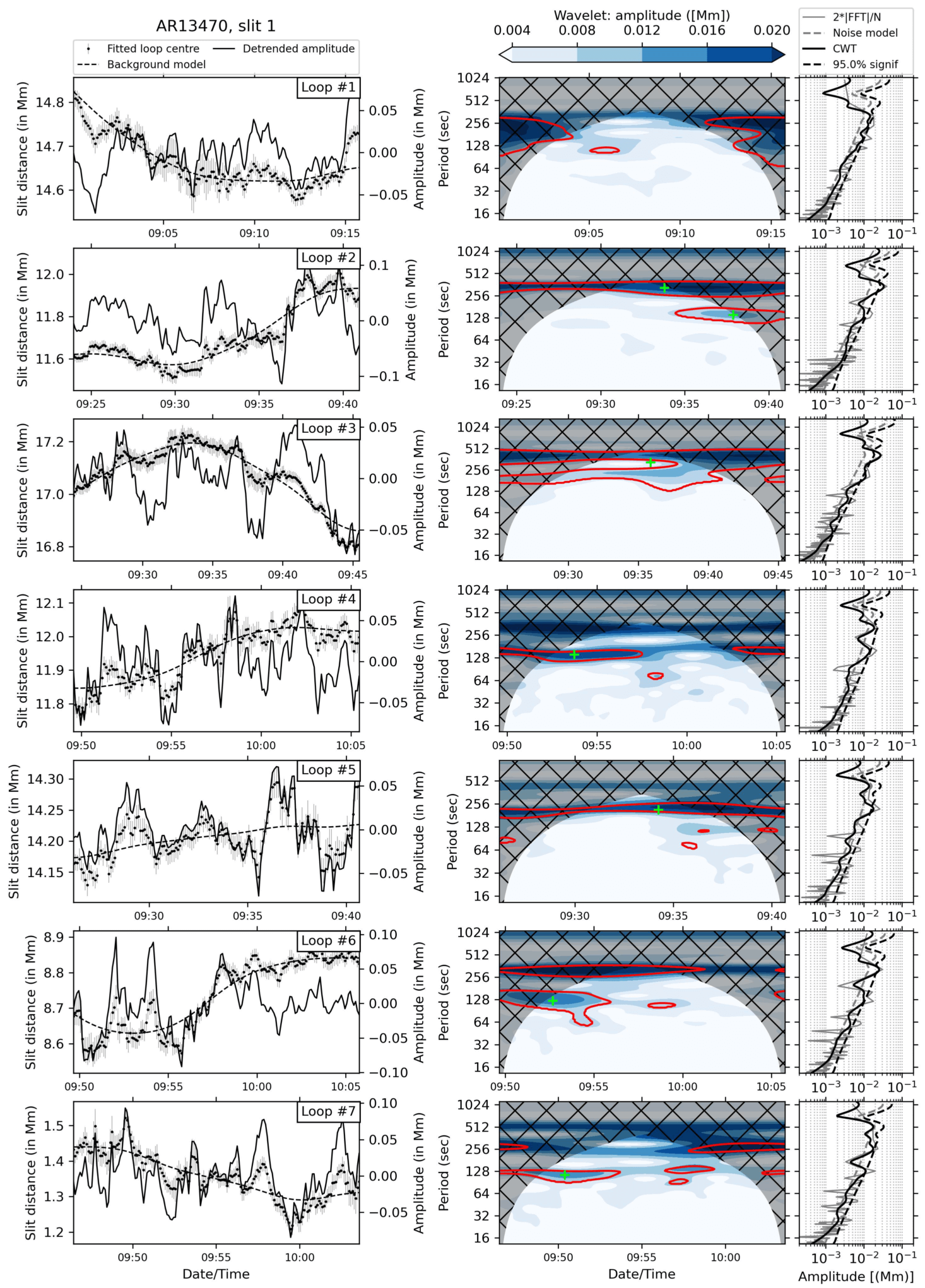}
	\caption{Coronal oscillation analysis for the plage/enhanced-network coronal loops fitted in slit 1 of AR13470. First column: time series for the loop fitted centres (dots) including uncertainties (light grey vertical bars) and background profile (dashed line) on the left y-axis, and final detrended oscillation amplitudes on the right y-axis (solid line). Second column: the (Morlet) wavelet amplitude of the detrended oscillations with a $95\%$ confidence interval (red contours). Third column: global wavelet power averaged over the whole time series (solid black line), red noise model (dashed grey line) and corresponding significant spectrum at $95\%$ (dashed black line). The power spectra calculated from a Fourier analysis is also plotted for comparison (solid grey line).}
	\label{fig:2023-10-19_plage-enhanced_allwavelets_slit1}
\end{figure*}

The short coronal loops in Fig.~\ref{fig:2023-10-19_plage-enhanced_Imaps} are delicate to analyse since they appear as multiple fine strands that often overlap in the time-distance maps. Another challenge arises from the potential contamination by lower atmospheric structures such as dynamic fibrils or spicules. Nevertheless, we could identify and fit several relatively bright and isolated coronal loop structures that are depicted by white solid lines in Fig.~\ref{fig:2023-10-19_plage-enhanced_Imaps}. A multi-Gaussian fitting method was employed to track the centre locations of these loops labelled loop \#1-\#7. As described in Sect.~\ref{sec:data-method_corona} and in Appendix~\ref{sec:appendix_method_corona}, the time series are then detrended to reveal any oscillatory signature with period below 10\,min. The detrended profiles for the oscillation displacement amplitude and the results of the wavelet transform are shown in Fig.~\ref{fig:2023-10-19_plage-enhanced_allwavelets_slit1}. At first look most of the fitted loops show oscillations around $\approx 300\ \rm{s}$ period which can be a sign of the leakage of the five-minute photospheric p-modes in the corona. This has already been shown to occur in short transition-region or low-coronal loops \citep{Gao2023a,Gao2023b}. Contamination from lower atmospheric signals is also not excluded at this stage. Other periods that more likely belong to the kink mode of interest for this study are discussed in the next paragraphs.

The loop sections labelled 2 and 4 constitute the same loop but at different times. They show similar periods at $\approx 128\ \rm{s}$ compatible with the kink-mode periods reported in recent EUI/HRI studies \citep[see e.g.][Fig.~5 and references therein]{Shrivastav2024}. Assuming a simple semi-circular geometry, an estimate of our loops length is about 35-70\,Mm. This lies in the range of loop lengths studied in \citet{Gao2022} who also detected kink-mode oscillations with similar periods. Higher harmonics have also been detected in past observations \citep{Verwichte2004,Doorsselaere2007,White2012,Pascoe2016,Duckenfield2018}. However it is unlikely that the oscillation periods detected here correspond to the second harmonic of the kink mode, because it has a node at the loop apex around which we traced slit 1. The third harmonic has an anti-node at the loop apex as does the fundamental, but the third harmonic has been detected mostly in large-amplitude oscillations associated with impulsive flare events.

Near co-temporal fits of adjacent loops have also been performed to check for any shared oscillation property as expected from the global behaviour of the kink mode. While the aforementioned $\approx 300\ \rm{s}$ period is present over the whole interval in both loop 4 and 6, the $\approx 128\ \rm{s}$ period (which is most likely associated to the kink mode) is shared with loop 6 only for the first half of the time interval. In a similar way, only loops 3 and 5 share common oscillation periods at $\approx 180-200\ \rm{s}$. This is again a reasonable period expected for the kink mode.

Finally, most of the fitted oscillations have displacement amplitudes within the 0.05-0.1\,Mm range which agrees with the statistical study from \citet{Gao2022} performed on similar loop lengths. However, these amplitudes are on the lower end of the distribution compared to the small-scale active region loops analysed by \citet{Li2023}, where the loop centres and edges were manually fitted.

\subsection{AR13470: plage/plage loops}
\label{sec:results_corona_plage-plage}

Solar plages with their high magnetic flux concentration are commonly observed in active regions. Coronal loops connecting to such regions probably constitute most of the observed loops. In this section we investigate a bundle of coronal loops with lengths $\approx 200-300\ \rm{Mm}$ that connect two plage regions in AR13470. One footpoint is anchored in region D of Fig.~\ref{fig:Oct19_context}, while the other footpoint is anchored in a plage of opposite polarity located outside of the SST/CRISP FOV and south-east of region D (see Fig.~\ref{fig:obs_context}). Slit 2 is placed near their apex.

The time-distance map for slit 2 is shown in Fig.~\ref{fig:2023-10-19_plage-plage_allwavelets_slit2a} in Appendix \ref{sec:appendix_coronal_results_all}. Oscillation patterns can be seen almost everywhere. However, the fitting process has been delicate here due to overlap of multiple loops along the line of sight, as well as apparent merging and splitting of the loop strands. An example of an apparent splitting can be seen in Fig.~\ref{fig:2023-10-19_plage-plage_allwavelets_slit2a} from time 09:40~UT at 15\,Mm from the edge of the slit, with a strong increase in intensity. Such event could be related to partial reconnection between entangled loop strands as seen in \citep{Antolin2021}. Given all this, an attempt was still made to fit the centre of two apparent loop strands which can be identified by the solid white lines in the top panel of Fig.~\ref{fig:2023-10-19_plage-plage_allwavelets_slit2a}.

The wavelet and Fourier analyses that result from these two fits are shown in the lower panels of Fig.~\ref{fig:2023-10-19_plage-plage_allwavelets_slit2a}. Two main oscillation patterns can be seen with a long $\approx 400-500\ \rm{s}$ and shorter $\approx 200-300\ \rm{s}$ period. The signal corresponding to loop 2 is unfortunately too short to allow the recovery of the longer period, and loop 1 shows only a few oscillation cycles of the short period. We suspect that both periods are at play in both loops as they are part of the same bundle. The long $\approx 400-600\ \rm{s}$ period and its related displacement amplitudes of $0.1-0.2\ \rm{Mm}$ agree with the values typically observed for the fundamental kink mode for loop lengths $\approx 200-300\ \rm{Mm}$ \citep[see e.g.][]{Anfinogentov2015}. The shorter $\approx 200-300\ \rm{s}$ period which is about a half or even a third of the long period could be a signature of the second or third harmonic. Again, the second harmonic is very unlikely to be detected here since the loop apex should behave as a node with no displacement, and the third harmonic is rarely observed in decay-less kink-mode oscillations. Another possibility is that the shorter $\approx 200-300\ \rm{s}$ period corresponds to remnants of the photospheric five-minute p-modes as mentioned earlier. A strategy to confirm this would be to trace other slits at the expected anti-node locations for the third harmonic (at $1/6$ and $5/6$ along the loop axis), and check for any phase coherence in the oscillations at that period. This would be non-trivial, as it is difficult to trace the entire loop in our observations due to the line-of-sight overlap with other features. We will retain here that the loop bundle investigated here (or at least loop 1) shows signatures of oscillations that are compatible with the fundamental kink mode.

\subsection{AR13470: pore-pore coronal loops}
\label{sec:results_corona_pore-pore}

We examine in this section a very long $\approx 400\ \rm{Mm}$ loop bundle connecting the negative and positive polarities of AR13470, namely the pores located within region C of Fig.~\ref{fig:Oct19_context} and the pores located much further south (see Fig.~ \ref{fig:obs_context}). 

The results of the coronal oscillation analysis along slit 3 are shown in Fig.~\ref{fig:2023-10-19_pore-pore_allwavelets_slit3a} in Appendix \ref{sec:appendix_coronal_results_all}. Oscillatory patterns are more difficult to detect here because the long loops appear more diffuse, dimmer and exhibit less of a thin strand-like appearance. Nonetheless a fit was made to one persistent loop at the middle of the slit (loop 1), with the aim of having a long enough time series to capture the period of the fundamental kink mode. In such long loops, the period associated to the kink mode is expected to be longer than the other loops of the same active region analysed in section \ref{sec:results_corona_plage-enhanced} and \ref{sec:results_corona_plage-plage}. In the wavelet analysis, some weak power can be seen indeed at periods $\approx$~10-13\,min which would correspond to the fundamental kink mode for such long loops \citep[see e.g.][]{Anfinogentov2015}. The displacement amplitudes are about the same order of magnitude as for the shorter loops studied in section \ref{sec:results_corona_plage-plage}. Since the displacement amplitude of the kink mode tends to scale with the loop length \citep[see e.g.][]{Zhong2022a}, that gives some hint that potentially less power was provided to excite these long loops.

\subsection{AR13468: coronal loops connecting to a sunspot}
\label{sec:results_corona_sunspot}

A bundle of coronal loops, or at least their leg, can be seen in EUI/HRI to connect in the vicinity of the AR13468 sunspot observed by SST/CRISP and analysed in section \ref{sec:results_photosphere_sunspot} (see Fig.~\ref{fig:Oct20_context}). More specifically they seem to connect to its penumbra where small-scale photospheric magnetic elements are continuously moving out from the sunspot (see section \ref{sec:results_photosphere_sunspot}). The fact that only the leg of these loops is visible from the EUI/HRI perspective will limit the diagnostic that can be made from the oscillation properties since their length cannot be estimated.

We focus on the thinner strands that are visible in the lower part of the sunspot (see Fig.~ \ref{fig:Oct20_context}). We first check whether these loops are oscillating or not by cutting the slits 4a-b-c across them, for which the EUI/HRI corresponding time-distance maps are shown in Figs.~ \ref{fig:2023-10-20_sunspot_allwavelets_slit2}-\ref{fig:2023-10-20_sunspot_allwavelets_slit3} in Appendix \ref{sec:appendix_coronal_results_all}. We used the same procedure as above to extract oscillatory properties. For all first three fitted loops shown in Fig.~\ref{fig:2023-10-20_sunspot_allwavelets_slit2} there is significant power at $\approx$200-300\,s. Interestingly loop 4 seems to behave differently with some oscillations at a longer period of $\approx$10\,min. Since loops 1-3 seem to be part of the same bundle as loop 4, one would expect similar loop lengths and hence similar kink-mode oscillation periods. A possible interpretation is that the oscillation periods of loops 1-3 at $\approx$200-300\,s do not belong to the kink mode but may simply be remnant of photospheric oscillations. They could indeed result from the often suspected leakage of the ubiquitous five-minute photospheric p-modes into the chromosphere, solar corona \citep[see e.g.][]{Morton2016,Morton2019} and beyond \citep{Huang2024arXiv}. However there is a higher chance that the $\approx$3-5\,min coronal oscillations detected here are related to the innate nature of the sunspot itself. It is not in the scope of this paper to study these sunspot oscillations in SST/CRISP. However, we will mention that we could see traces of them in the sunspot core in both the SST/CRISP and EUI/HRI time series, and that they have received an extended coverage in the literature \citep[see][for a review on that topic]{KhomenkoCollados2015review}. Indeed three-minute oscillations are systematically observed in sunspot umbrae \citep[commonly known as umbral flashes, see ][]{Rouppe2003,Kobanov2008}, and they have been detected to leak into the corona above by following connected coronal loops \cite[see e.g.][]{Sych2009,Jess2012}. On the other hand five-minute oscillations have been detected in sunspot penumbrae as running penumbral waves \citep[see e.g.][]{Rouppe2003,Kobanov2008}. Sunspot penumbrae being mostly made of highly horizontal magnetic field \citep{Title1993}, running penumbral waves could possibly apply the required transverse kinks that propagate along the connected loops and are detected further out in the corona. It is not straightforward though how both three- and five-minutes sunspot oscillations may convert into transverse oscillations in the corona. EUI/HRI could be crucial in making that connection thanks to its unprecedented high-resolution and high sensitivity to lower atmospheric features, that is left for future studies.

As a further check we repeated the analysis at two different locations along the loops axis with slit 4a and 4c, and both the short $\approx$200-300\,s and long $\approx$10\,min oscillation periods were found (see Fig.~\ref{fig:2023-10-20_sunspot_allwavelets_slit1} and \ref{fig:2023-10-20_sunspot_allwavelets_slit3} in Appendix \ref{sec:appendix_coronal_results_all}). Given that the coronal loops studied here are partially visible, we will just conclude that the long $\approx$10\,min period would be compatible with the expected period of the standing kink mode in very long ($>$500\,Mm) loops, that would then appear as open when seen from above by EUI/HRI. Furthermore the displacement amplitudes vary between $0.05$ and $0.1$\,Mm similarly to the loops studied in the other sections above. However, the displacement amplitudes have here been measured close to the loops footpoint and would be expected to be larger if measured close to the loop apex.


\section{Results: the photosphere-corona connection}
\label{sec:results_photosphere-corona}

\subsection{Summary and qualitative interpretation}
\label{sec:results_photosphere-corona_summary}

Most of the analysed coronal loops exhibit double-period oscillation signatures, with one of the periods being compatible with the fundamental kink mode. The secondary detected oscillations have a common period around three to five minutes regardless of the loop length, and as such are suspected to be coronal counterparts of the photospheric p-modes that occur on a global scale. We now discuss the detected coronal oscillations that can be related with the fundamental kink mode. It is important to note that not all of the coronal loops visible in EUI/HRI (and even those that have been fitted) exhibit clear kink-mode oscillations. Albeit these are often considered ubiquitous in the solar corona, there seem to be conditions that are not favourable for their development. On the other hand, the analysis of the photospheric regions at the base of these coronal loops reveals that the transverse motions can vary in strength from one region to another. This suggests a connection between the kink-mode oscillations in coronal loops and the dynamics at the photosphere.

Among the loops that showed the weakest oscillatory behaviour are the long ($\approx 400\ \rm{Mm}$) loops analysed in slit 3 of AR13470 (see Sect.~\ref{sec:results_corona_pore-pore}). These long loops connect into photospheric regions where the magnetic field flux concentration is the highest and where pores are present (see region C in Fig.~\ref{fig:Oct19_context}). Among all the photospheric regions, the region C has the weakest photospheric motions, even compared to the more regular plage regions B and D. This suggests that less energy than usual was available to trigger the kink-mode oscillations in these long loops, or that perhaps the excitation mechanism is less efficient there.

On the other side, the short ($\lesssim 70\ \rm{Mm}$) loops located in the core of AR13470 and investigated in slit 1 have shown a lot more dynamics with larger amplitude oscillatory signals overall (see Sect.~\ref{sec:results_corona_plage-enhanced}). Short coronal loops in the core of active regions are often observed to be highly dynamic \citep[see e.g.][]{Li2023}. This is also a place where the photosphere is often changing due to the emergence of magnetic flux. In the present case of AR13470 we do see some flux emergence close to its core (see Sect.~\ref{sec:results_photosphere_Oct19}). However the short loops tracked in slit 1 are likely connecting too far south of that region (see region B in Fig.~\ref{fig:Oct19_context}), which is a regular plage with no apparent emerging magnetic flux. The other footpoints of these loops are located in a different type of photospheric region though, namely the enhanced network (region A) that shows stronger motions on long timescales (i.e. higher $v_h$). Therefore the base of the short loops connecting to region A are likely more affected by the neighbouring convection, and as a consequence the photospheric driving should be enhanced there. That would at least partially explain why such loops show stronger kink-mode oscillations.

A final important case to discuss is the coronal loops that are anchored in the sunspot of AR13468. Even though most of the oscillations within these loops were attributed to the photospheric oscillations produced by the sunspot itself (i.e. the three- and five-minute oscillations, see section \ref{sec:results_corona_sunspot}), potential traces of the kink mode could be detected as well. Given that such long ($>500\ \rm{Mm}$) loops would eventually connect back to the solar surface, the period of $\approx$10\,min measured would be compatible with the fundamental kink mode. Unfortunately, the analysis was limited by the short portion of the loop legs visible in EUI/HRI. As a consequence, if these oscillations really belong to the fundamental kink mode then the displacement amplitudes that we measured do not reflect the reality and are greatly under-estimated. Controversially the loop footpoints appear to be steadily anchored in the sunspot and hence a photospheric driving of kink-mode oscillations in such loops is often questioned. However, the medium around or beneath these loop footpoints is far from being static. We showed in section \ref{sec:results_photosphere_sunspot} that there are systematic motions in the sunspot penumbra where the coronal loops are anchored, and that these motions are sustained over a hour at least. Such motions would appear as a quasi-steady or low-frequency broadband photospheric driver and are strong enough ($\approx 0.4\ \rm{km/s}$) to trigger the development of kink-mode oscillations (see e.g. \citet{Karampelas2020}). We conclude that while sunspots are often seen as one of the most static features from the perspective of coronal (loop) EUV observations, they remain highly dynamic in the photosphere and chromosphere. Consequently, sunspots are favourable environments for the development of coronal oscillations, including kink modes.

\subsection{Quantification of the photosphere-corona connection}
\label{sec:results_photosphere-corona_quantification}

In order to further establish the role of the photosphere in the excitation of coronal loop kink oscillations, a quantification is needed of both photospheric and coronal diagnostics. The photospheric dynamics was quantified with three parameters $v_h$, $a_1$ and $a_2$ that allow us to quantify the photospheric driving (see section \ref{sec:results_photosphere} and table \ref{tab:results_photosphere}).

\begin{figure}
    \centering
    \includegraphics[width=1.\columnwidth]{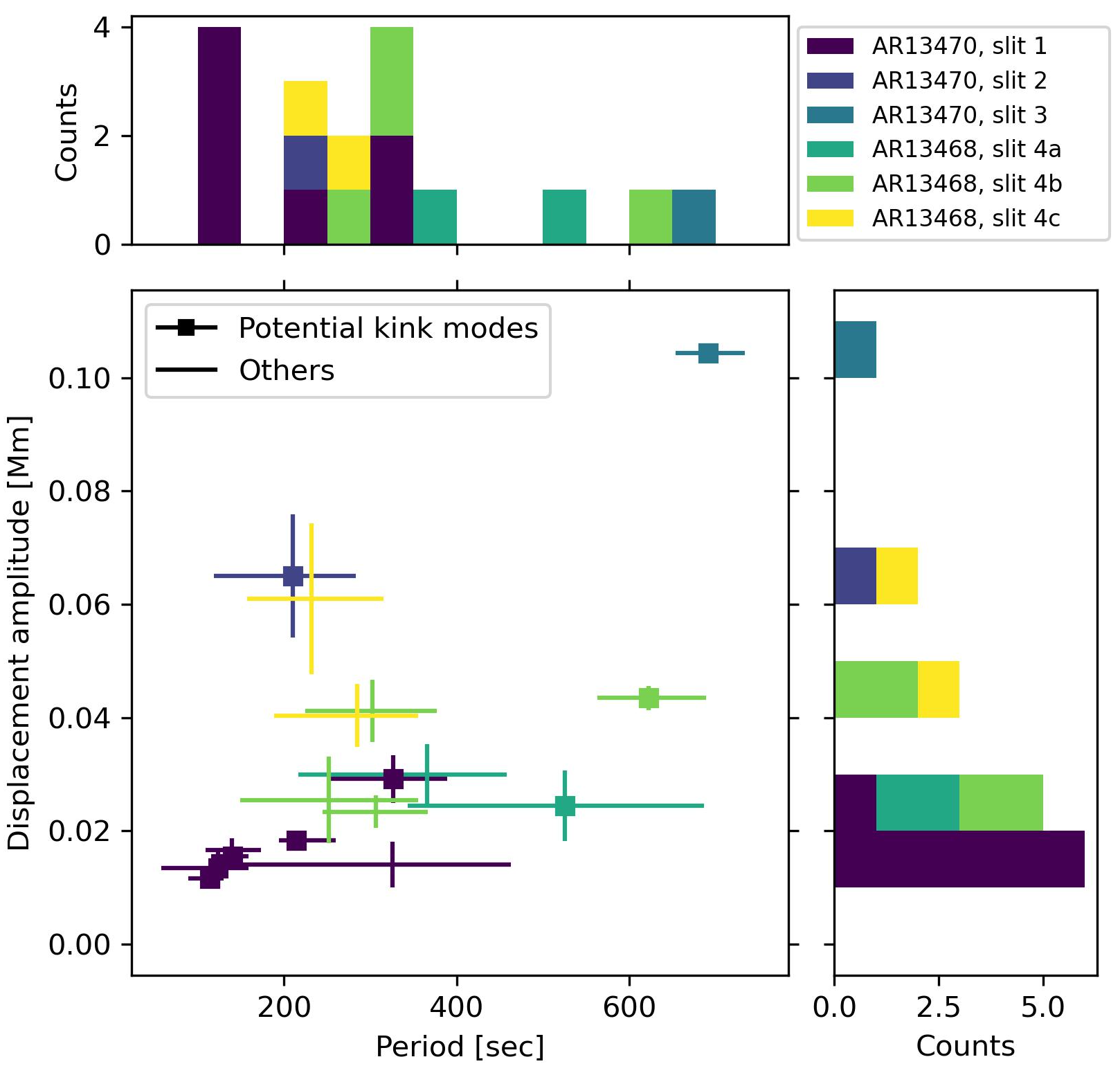}
	\caption{Distribution of oscillation amplitudes versus period from the clustered wavelet results. }
	\label{fig:all_COR_results}
\end{figure}

Although the active region loops observed by EUI/HRI and considered here are not ideal for clear oscillation identification and quantification, we will still establish the methodology and show its potential for future works. The extensive results given by the wavelet transforms presented in section \ref{sec:results_corona} can be reduced in clusters where averaged values for the oscillation period and displacement amplitude can be extracted by applying some criteria to keep only the significant signals (see Appendix \ref{sec:appendix_method_corona} for more details). The results of this clustering are shown by the green crosses in the wavelet plots, and are summarised in Fig.~\ref{fig:all_COR_results}. Each cluster was labelled as a potential kink mode (square markers) following the discussion in section \ref{sec:results_corona}. A large number of coronal oscillations can be found at around $\approx$3-5\,min which were mostly identified as oscillations of photospheric origin. On the other hand, detections associated with kink modes are distributed over a broad range of periods, which depend on the loop length as well as on other properties such as the plasma density and magnetic field strength. As expected from the fundamental kink mode, the displacement amplitudes that were measured close to the loop apex (except for the loops of AR13468) tend to scale with the period. In order to compare the kink-mode properties between loops of different lengths it is more adequate to use the velocity amplitude defined as $2\pi \times Amplitude/Period$ \citep[see e.g.][]{Li2023,Shrivastav2024}. Variations in the velocity amplitude would then potentially result from a different photospheric driver instead of a difference in the loop geometry and coronal conditions.

\begin{figure*}
    \centering
    \includegraphics[width=2.\columnwidth]{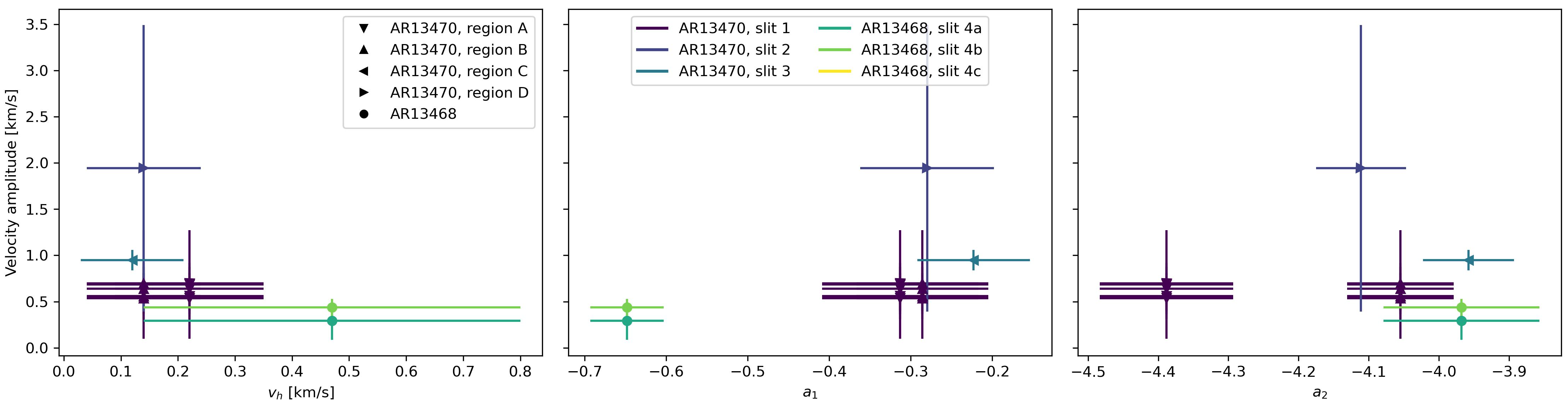}
	\caption{Scatter plots of the coronal oscillation velocity amplitude for the potential identified kink modes, versus the motion properties ($v_h$, $a_1$, $a_2$) of the corresponding photospheric regions.}
	\label{fig:all_COR+PHO_results}
\end{figure*}

The oscillation velocity amplitude for the potential identified kink modes only are plotted in Fig.~\ref{fig:all_COR+PHO_results} against the photospheric driving parameters derived in section \ref{sec:results_photosphere}. The results are colour-coded similarly as in Fig.~\ref{fig:all_COR_results} with markers that indicate the photospheric region(s) where the loops connect to. This allows to study cases where oscillating coronal loops connect to photospheric regions of different type and dynamics, such as in the case of the loops fitted in slit 1 of AR13470 (dark blue colour) that connect both to a plage (region B, up-oriented triangle) and an enhanced network (region A, down-oriented triangle). 

\subsection{Discussion}
\label{sec:results_photosphere-corona_discussion}

A first observation is that the oscillation velocities vary among the loop bundles and seem to depend on their connectivity down to the photosphere. Coronal loops would be expected to react differently depending on the strength and frequency distribution of their photospheric driver. As discussed in the introduction there is suspicion that coronal loops manifest both self- and forced-oscillatory behaviours. The latter would naturally arise due to the overlap in frequency of the kink modes with some photospheric oscillations, such as the global p-modes and sunspot oscillations. Such coupling has already been demonstrated in simulations \citep{Ballai2008,Gao2023b} but would be the most efficient only in coronal loops with a compatible resonant frequency, which in our case would correspond to the loops fitted in slit 1 and 2 of AR13470 (dark and medium-dark blue colours). Furthermore the convection and transport of magnetic flux at timescales comparable with that of the kink mode may also trigger the forced-oscillatory behaviour in the compatible loops. 

However the self-oscillatory excitation mechanism is believed to be the most systematic because it agrees with most of the properties of observed kink modes \citep{Nakariakov2016a}. The two forces involved in the self-oscillatory process would correspond to the magnetic friction between neighbouring shearing quasi-parallel magnetic field lines and the magnetic tension force. A key aspect is that this excitation process is the most efficient when the friction between the system and the exterior occurs at some distance from the tied points \citep{McLennan2008}. Practically speaking, the chromospheric and/or transition region heights may hence be the locations where this stick-slip interaction is the most efficient at driving the coronal loop footpoints. In the ideal case of \citet{Helmholtz1954} where the sticking phase is complete, the system oscillates exactly at the speed of the driver. Translated to the case of oscillating coronal loops, the velocity amplitude of the detected kink oscillations would then be expected to follow closely the driving velocity of the loop footpoints.

In practical terms, this self-oscillatory excitation mechanism would manifest as a positive correlation between the kink-mode velocity amplitude and the quasi-steady component $v_h$ of the photospheric driving parameters derived in this study. Some correlation would also be expected with the low-frequency slope of the broadband component $a_1$, a photospheric driving parameter that corresponds to timescales longer than the kink-mode period for most of the studied coronal loops. The type of correlation to be expected (e.g. linear or not) will need future dedicated theoretical and numerical work \citep[e.g. following][]{Nakariakov2016a,Nakariakov2022}. For now we can just state that there is no obvious correlation(s) between the parameters shown in Fig.~\ref{fig:all_COR+PHO_results}, however, more observational samples would be necessary before drawing a conclusion. It is worth to point out that the kink-mode velocity amplitudes for AR13468 (light and dark green dots) should in principle be much higher as discussed earlier, and hence would better agree with a positive correlation in the $v_h$ graph. The relationship between the coronal and photospheric parameters could also be more complex to interpret due to the forced-oscillatory behaviour that manifests in addition to the self-oscillatory one. In both cases though, the excitation is believed to be driven at the loop footpoint down to the photosphere and chromosphere. Recent numerical studies have shown the ability of the forced-oscillatory behaviour to excite decay-less kink oscillations in coronal loops, because of the high potential of broadband drivers to excite (or force) not only the fundamental but also the harmonic frequencies of the kink mode \citep{Karampelas2024,Karampelas2024b}. The relative contributions of the forced- and self-oscillatory mechanisms in the excitation of decay-less kink oscillations remain to be quantified. For instance, the damping pattern of the oscillation envelope can be used in observations to discriminate between the two excitation mechanisms \citep{Nakariakov2024}.

We presented here the methodology to combine photospheric and coronal observations in a quantitative and meaningful manner for the study of kink-mode oscillations in coronal loops and their potential excitation mechanism. This pilot study serves as a baseline for future works that investigate the photosphere-corona connection further.


\section{Limitations and outlook}
\label{sec:discussion}

Important limitations come with the usage of the LCT method to derive photospheric motions. For instance, extensive benchmarking studies have pointed out the difficulty in comparing the LCT outputs to actual velocity flows in test simulations of the solar granulation \citep[see e.g.][]{Verma2013,YellesChaouche2014,Louis2015}. LCT is efficient at retrieving relative displacements from contrast variations between successive intensity images (optical flows), but those are not necessarily associated with actual plasma flows. Furthermore, the LCT method has proven to be an efficient and fast method that allows to recover at least the morphology of photospheric horizontal motions despite a tendency to underestimate the velocity magnitude. Since the LCT outputs can be heavily influenced by the size of the correlation window ($fwhm$) and the temporal cadence ($dt$), we systematically repeated our analysis on a set of eight different $fwhm$-$dt$ pair-parameters for which all results are given in table \ref{tab:results_photosphere}. The particular pair of $fwhm=600\ \rm{km}$-$dt=54\ \rm{s}$ was found to be the most optimal based on simulations of the solar photosphere (see Appendix~\ref{sec:appendix_method_photosphere} for more details).

Other methods have also been suggested recently giving better results overall such as the Fourier-LCT \citep[FLCT;][]{Fisher2008}, coherent structure tracking \citep[CST;][]{Roudier1999,Rieutord2007} or deep-learning based tracking \citep[DeepVel;][]{AsensioRamos2017}. Our aim here was not to derive actual flows at the photosphere, but to track and follow apparent motions of specific magnetic elements at the photosphere with high contrast in the continuum intensity. Therefore we claim that the LCT method should be sufficient to estimate the amount of photospheric driving beneath oscillating coronal loops. These results can already be tested in simulations and, hopefully, lead to more insights on the excitation mechanism of oscillating coronal loops. A future study could test the other aforementioned tracking methods to see how it may improve the precision of our results, however the focus should first be on extending the pool of observational datasets to achieve more statistically reliable results. Finally, we stress out that having longer time series of the photospheric motions may also improve the accuracy of the low-frequency part of the Fourier analyses, that is a challenge to be addressed in future coordinated observations.

While the methodology for the analysis of coronal oscillations is already quite established and has been extensively tested in the past, it also suffers some weaknesses. 

A first difficulty comes with the co-alignment of the coronal images together at a sub-pixel precision. We showed that meaningful oscillating signal could already be extracted even without the use of such sophisticated techniques. However our analysis could have revealed more kink-mode signatures if the noise induced by the spacecraft jittering had been better corrected. Therefore our coronal oscillation results likely represent a limited portion only of the actual kink modes contained in the observed coronal loops.

Our estimates of the oscillation displacement amplitude (and in turn velocity amplitude) can be affected by several factors including the loop inclination with respect to the line-of-sight, the spacecraft jittering and the wavelet transform. The latter tends to flatten out the response over the frequency dimension as well as to underestimate the oscillation amplitude compared to the classical Fourier approach. The choice of the mother wavelet can be important in that regards. We chose nonetheless the classical Morlet function to be consistent with past studies of decay-less kink oscillations. Both the jittering and wavelet side effects are presumed to act similarly in all studied loops. As a consequence, they should not alter our conclusions since we focused on the relative variations between loop bundles and not on the absolute values.

The choice of the detrending profiles to obtain the kink-mode amplitudes was also a critical part of the methodology. We systematically checked that the high-pass Fourier filters used did not alter too much the signal of interest, but there might still be a minor impact on the deduced oscillation amplitudes close to the filter cutoff frequency (see Appendix~\ref{sec:appendix_method_corona}).

The SST offers a valuable set of chromospheric observations, aiding detailed studies of the photosphere-corona connection. While chromospheric spectral lines can provide precious information on wave propagation and motions, the chromosphere also adds a lot more of complexity in terms of dynamics and small-scale structures. Making the connection between the coronal loop footpoints seen in EUI/HRI and the chromosphere seen by SST/CRISP can be insightful but has inherent additional challenges that would have had to be overcome, such as high-precision cross-instrument alignment and issues related with the different perspective angles. Inference of the magnetic topology in the chromosphere may also be crucial towards achieving this goal. However, that was beyond the scope of this study and is left for future works. The European Solar Telescope \citep[EST;][]{Quinteronoda2022} and its focused design on highly-sensitive spectropolarimetry in the chromosphere will certainly become crucial to make that photosphere-corona connection.

\section{Conclusion}
\label{sec:conclusions}

The driving and excitation mechanism of sustained kink-mode oscillations in coronal loops remain a mystery. Coronal loop footpoints often appear as "static" in (E)UV images and as a result the possibility of photospheric driving of kink-mode oscillations in coronal loops has often been rejected. In the detailed introduction, we showed that such a photospheric driving has nonetheless received a lot of support from both simulations and observations in recent years. We further investigated the photosphere-corona connection in this context by exploiting an unique set of dedicated high-resolution observations that were taken in coordination from both space with SolO and the ground with the SST in October 2023.

The first part of this study highlights the dynamism of the photosphere as seen in high-resolution observations by the CRISP instrument at the SST. We used a local correlation tracking technique to estimate horizontal motions in specific sub-regions where overlying coronal loops in EUI/HRI were observed to connect. We showed that these motions vary from one photospheric region to another and increase overall in strength going from pore, plage, enhanced-network to sunspot regions. These motions can be quantified with a quasi-steady component and a broadband component, where the latter can be further divided into a low-frequency and high-frequency component. Each component can be associated to their respective scale in the photosphere, spanning from large (super-granulation) to small (granulation) scales and even below. Our results show counter-intuitively that coronal loops anchored steadily in sunspot surroundings would be the most affected by photospheric driving. A photospheric driving of oscillating coronal loops that connect to pore or plage regions is not to be excluded either. While being greatly reduced in such regions, the quasi-steady and low-frequency components of the photospheric motions are still non-negligible, and the high-frequency part remains mostly unaffected. 

If kink-mode oscillations are indeed driven by the lower atmosphere, a difference in the properties of these oscillations is then expected depending on the loop connectivity into the photosphere (and chromosphere). We investigated such possibility by analysing coronal loops in EUI/HRI that connect to the photospheric regions analysed in the first part. Traces of the fundamental kink mode could be found in several of these coronal loops. Most of the studied coronal loops also showed secondary oscillation patterns at around 3-5\,min that seem to be of photospheric/chromospheric origin, in agreement with the global p-modes and sunspot oscillations. 

We concluded this work by combining the photospheric and coronal results together. Coronal loops may act as both forced and self oscillators in response to photospheric driving. Although the former could explain the observed coupling of the 3-5\,min photospheric oscillations with the kink mode in the corona, the main contribution to the kink-mode excitation is believed to manifest as a self-oscillatory behaviour. Indeed coronal loops as self oscillators have the ability to convert driving motions that are seemingly steady on relatively long timescales (with respect to the kink-mode period) into self-sustained resonant kink-mode oscillations. An evidence of such behaviour would be a correlation between the velocity amplitude of the kink-mode oscillations and the velocity of the (photospheric) driver at zero ($v_h$) and low ($a_1$) frequency.

If no proof of the self-oscillatory behaviour can be established yet with the limited set of observations investigated here, there are compelling signs though that the dynamics in the photosphere (and chromosphere) is intimately intertwined with the excitation of kink oscillations in coronal loops. In that sense this work may serve as a pilot study and baseline for future works that aim to investigate the photosphere-corona connection further, and will be continued when new dedicated and coordinated observations will be acquired. In the meantime, simulations can be crucial to investigate in more details the self-oscillatory excitation mechanism and its associated stick-slip interaction in realistic magnetic field configurations. To this end, the observation-derived parameters provided in this work for the photospheric quasi-steady and broadband driving can be critical at better constraining existing coronal loop models that investigate coronal oscillations and their counterparts as coronal heating. 

\begin{acknowledgements}
We acknowledge the anonymous referee for the valuable comments and suggestions.

NP and PK acknowledge funding from the Research Council of Norway, project no. 324523. This research was also supported by the Research Council of Norway through its Centres of Excellence scheme, project no. 262622, and through project no. 325491. 
This project has received funding from the Swedish Research
Council (2021-05613) and the Swedish National Space Agency (2021-00116). 
The Swedish 1-m Solar Telescope is operated on the island of La Palma by the Institute for Solar Physics of Stockholm University in the Spanish Observatorio del Roque de los Muchachos of the Instituto de Astrofísica de Canarias. The Swedish 1-m Solar Telescope, SST, is co-funded by the Swedish Research Council as a national research infrastructure (registration number 4.3-2021-00169).

Solar Orbiter (SolO) is a space mission of international collaboration between ESA and NASA, operated by ESA. We are grateful to Dr. Janvier M. and Dr. Hayes L. who contributed to the planning of the SolO observation datasets used in this study and to the coordination with ground-based facilities.

We are also grateful to the ESA SOC and MOC teams for their support. The EUI instrument was built by CSL, IAS, MPS, MSSL/UCL, PMOD/WRC, ROB, LCF/IO with funding from the Belgian Federal Science Policy Office (BELSPO/PRODEX PEA 4000134088); the Centre National d’Etudes Spatiales (CNES); the UK Space Agency (UKSA); the Bundesministerium für Wirtschaft und Energie (BMWi) through the Deutsches Zentrum für Luft- und Raumfahrt (DLR); and the Swiss Space Office (SSO). The German contribution to SO/PHI is funded by the BMWi through DLR and by MPG central funds. The Spanish contribution is funded by AEI/MCIN/10.13039/501100011033/ and European Union “NextGenerationEU”/PRTR” (RTI2018-096886-C5, PID2021-125325OB-C5, PCI2022-135009-2, PCI2022-135029-2) and ERDF “A way of making Europe”; “Center of Excellence Severo Ochoa” awards to IAA-CSIC (SEV-2017-0709, CEX2021-001131-S); and a Ramón y Cajal fellowship awarded to DOS. The French contribution is funded by CNES.

The authors have extensively used the open source \emph{JHelioviewer} software developed by ESA under the contract No. 4000107325/12/NL/AK - High Performance Distributed Solar Imaging and Processing System - and run at the Solar Influences Data Analysis Center (SIDC) of the Royal Observatory of Belgium (ROB). This paper has also made use of the NASA Astrophysics Data System (ADS) and the \emph{ImageJ} image processing tool developed by Wayne Rasband and contributors at the National Institutes of Health, USA.

\end{acknowledgements}

\bibliographystyle{aa}
\bibliography{Poirier2024}

\begin{appendix}

\section{The methodology in details, photospheric motions}
\label{sec:appendix_method_photosphere}

Motions in the photosphere are calculated from the SST/CRISP wide-band continuum intensity of \FeI~6173~\AA\ for the Oct-19 dataset and \Halpha~6563~\AA\ for the Oct-20 dataset. The time sequences are already co-aligned as part of the SSTRED reduction pipeline \citep{Lofdahl2021}. Due to unstable atmospheric seeing the SST/CRISP observations on Oct-20 were taken for \Halpha~6563~\AA\ only and in trigger mode, meaning that data was acquired only when the seeing quality was above a certain threshold. Additionally, blurry frames were removed when the contrast (absolute difference between the $1\%$ and $99\%$ percentile normalised by the median) of the SST/CRISP \Halpha~6563~\AA\ wide-band intensity over the full FOV was below a threshold of $0.6$. Both the $r_0$-trigger mode and post cleaning stage lead to a SST/CRISP dataset that is non-equidistant in time. The SST/CRISP datacube was finally re-interpolated in the temporal dimension (piecewise interpolation) to get a constant cadence of $27\ \rm{s}$.

Sub-FOVs of $20"$ size are selected for the motion analyses in order to keep track of memory usage, computational time and to make the visualisation of motions easier. Persistent scintillation from the photospheric p-modes is removed using a subsonic Fourier cone filter $\omega/k>7\ \rm{km/s}$. A LCT \emph{Python} algorithm\footnote{https://github.com/Hypnus1803/pyflowmaps} is then employed to calculate \emph{apparent} horizontal motions. Two important parameters must be specified to the LCT code, a size for the cross-correlation window ($fwhm$ for full-width-at-half-maximum) and the time delay $dt$ between the first and last time frames given. The LCT method has been extensively tested in the past and is known to be heavily influenced by these two parameters \citep[see e.g.][]{Verma2011,Verma2013}. Therefore we systematically tested with many pairs of LCT parameter values.

We used two reference radiative-MHD simulations of the solar photosphere to find the optimal set of LCT-parameters for this study, namely $\rm{CO^5BOLD}$ \citep{Wedemeyer2004,Freytag2008} in a pure non-magnetic solar granulation configuration \citep[see][and referenced therein]{Wedemeyer2009}, and Bifrost \citep{Gudiksen2011} in an enhanced-network setup \citep{Carlsson2016,Kohutova2021,Kohutova2023}. In a similar approach as \citep{Verma2013,YellesChaouche2014,Louis2015}, we spatially degraded and temporally averaged the simulations to match the observational specifications of SST/CRISP. To allow a fair comparison with the LCT outputs, the actual flow maps from the simulations were spatially smoothed with a Gaussian window of the same size as the one used in the LCT method. In both degraded $\rm{CO^5BOLD}$ and Bifrost simulations we saw a similar two-part power-law Fourier spectrum for both the LCT-derived velocities and the actual flow velocities. The three photospheric driving parameters $v_h$, $a_1$ and $a_2$ were derived from both the actual flow maps and the LCT-derived motions. The best agreement, in the context of the SST/CRISP observations analysed in this paper, was found for the LCT-parameter pair $fwhm=600\ \rm{km},\ dt=54\ \rm{s}$.

Once the horizontal motions are obtained, we track the propagation of a grid of corks in time by interpolating the motion maps at all new cork locations at each time step. That allows us to compute the temporal Fourier power spectrum for the horizontal velocity along each trajectory. When the photospheric magnetic field can be inferred from the SST/CRISP \FeI~6173~\AA spectro-polarimetric observations (i.e. for the Oct-19 dataset), we further select only the trajectories with starting locations that match a given threshold on the magnetic field magnitude. This last step allows to track specific magnetic elements at the photosphere that have enough magnetic flux to possibly interact with and/or form coronal loops such as high magnetic flux concentrations related to active region plages, small-scale bright magnetic elements that propagate within the network or migrate away from sunspots. 

\section{The methodology in details, coronal oscillations}
\label{sec:appendix_method_corona}

EUV images from EUI/HRI$\mhyphen174$ are first enhanced using the wavelet-based enhancing technique of \citet{Auchere2023}, based on changing the local contrast by levelling out intensities among small and large scales features. The aspect of the enhanced image can be controlled with free parameters among which the weights for the de-noising filter $w_{noise}$, the enhancement "strength" $h$ and the $\gamma$-stretch scaling of the final intensities. There is not a single nor best set of parameters as the final image is only subjective. For this paper we used a set of parameters ($h=0.7$, $\gamma=3.2$ and $w_{noise}=[10,6,2]$) different than usually used for EUI/HRI ($h=0.995$, $\gamma=2.4$ and $w_{noise}=[5,3,1]$). That was motivated here by the need of emphasising finer loop structures to ease the oscillation detection process at the cost of a less realistic visual aspect.

A critical step is to co-align the EUI/HRI time series in order to suppress none physical displacements from one frame to the next. The \emph{Sunpy} \emph{Python} ecosystem is exploited to re-project the images on a single reference coordinate frame using the \emph{reproject\_to} method \citep{Sunpy2020}. This allows to suppress most of the jittering in the pointing thanks to the good precision of the World Coordinate System (WCS) coordinates stored in the EUI/HRI level-2 \emph{.fits} files. Alternative methods based on cross-correlation over consecutive frames were also tested, but they did not significantly improve the co-alignment precision, or at least not consistently throughout the whole timeseries.

The methodology is then very similar to past studies of kink oscillations: extraction of intensity variations along an artificial slit across a loop bundle to create a time-distance plot, selection of particular loops, multi-Gaussian fitting of the loop centre, and analysis of the transverse oscillation amplitudes with Fourier/Wavelet-based approaches. A few points are distinct from past studies. The artificial slit is defined with a certain small width ($4"$ for EUI/HRI) within which intensities are spatially averaged to increase the signal-to-noise ratio. We used again the \emph{lmfit} optimisation \emph{Python} library that allows to fit multiple profiles at once including a background profile (linear) in addition to as many Gaussian profiles as needed. This feature is useful when loops overlap in the coronal images which happens even more often in EUI/HRI. A \citet{Powell1964}'s minimisation algorithm was used here rather than the classical least-square algorithm because of its robustness against noisy data \citep{Press2007}. Once the loop centres are fitted for each time frame, the obtained time series must be de-trended to obtain the amplitudes of the loop oscillations. The choice of the background profile to subtract becomes then critical. High-pass Fourier filters were used to get rid of most slow background variations. Two cutoff frequencies of 10\, and 20\,min were selected to emphasize the kink-mode oscillations in the studied short and long loops respectively. The cutoff frequencies were carefully tuned to remove most of the unwanted oscillation power while preserving as much as possible the oscillations of interest. For transparency all background profiles are shown along with the final results.

The usage of wavelet transform is motivated here due to the complex nature of the data, where quasi-periodic signatures are often observed with sometimes time-varying frequencies as well. Wavelet analyses are performed in 1-D on each de-trended signal using the \emph{ev-tools} \emph{Python} package based on \citep{Torrence1998,Verwichte2004} and taking a Morlet-6 wavelet mother. Fourier spectra are also computed with the \emph{numpy} \emph{Python} library for control. The significant power from the wavelet transform is taken within the $95\%$ confidence interval, that corresponds to $2\mhyphen\sigma$ above the (red-)noise level. The de-trended signals follow very closely a red-noise power-law ($f^{-2}$) distribution over frequency. We use the approach of \citet{Torrence1998} based on a red-noise spectrum to estimate the noise level.

In order to characterise the properties of the oscillations, we further reduce the wavelet results into clusters. Contours of significance are first labelled as individual clusters using the \emph{measure} module of the \emph{skimage} \emph{Python} library \citep{scikit-image}. Average oscillation properties are then computed over each cluster such as the mean period (defined as the weighted centroid) and mean amplitude. Some criteria are finally defined to select the most meaningful clusters. The weighted centroids must be outside of the cone-of-influence which defines the area within the wavelet transform that is affected by edge effects \citep[see][]{Torrence1998}. The clusters duration must be at least $1.5$ times the mean period. The cluster mean period must be larger than 90\,s to filter out the remaining high-frequency signals that likely not belong to the fundamental kink mode of the targeted loops. Finally the period variation within each cluster must be below $150\%$ of relative error. These criteria have been specifically tuned for the datasets investigated in this paper but can hopefully serve as a basis for more exhaustive future studies.

\section{Photospheric motion analyses: full report}
\label{sec:appendix_photospheric_results_all}
For transparency we collected in table \ref{tab:results_photosphere} all results from the photospheric motion analyses run for this study.

\begin{table*}[]
	\caption{Full report of the photospheric motion analyses}
	\label{tab:results_photosphere}
	\centering
	\tabcolsep=0.11cm
    \resizebox{1.7\columnwidth}{!}{%
	\begin{tabular}{ccccccc}
		\hline\hline
                             & $\Bar{v_h}$            & $a_1$             & $a_2$             & $f_c$          &         & LCT params \\
		   Region  & $\rm{mean}\pm\rm{std}\ (\rm{max})$ & $\rm{mean}\pm\rm{std}$ & $\rm{mean}\pm\rm{std}$ & $\rm{mean}\pm\rm{std}$ & ROI & fwhm, dt \\
                             & [km/s]                 &                   &                   & [mHz]          &         & [km, sec] \\ 
		  \hline\hline
        AR13468              & $0.47\pm 0.53\ (0.31)$ & $-0.355\pm 0.049$ & $-2.203\pm 0.138$ & $7.67\pm 0.89$ & Sub-FOV & $300, \ 27$ \\
        sunspot w/ moat flow & $0.42\pm 0.44\ (0.31)$ & $-0.401\pm 0.053$ & $-2.315\pm 0.095$ & $5.48\pm 0.56$ &         & $600, \ 27$ \\
                             & $0.36\pm 0.26\ (0.34)$ & $-0.661\pm 0.061$ & $-2.509\pm 0.151$ & $6.96\pm 0.97$ &         & $1200,\ 27$ \\
                             & $0.44\pm 0.45\ (0.31)$ & $-0.404\pm 0.038$ & $-3.375\pm 0.098$ & $6.45\pm 0.28$ &         & $300, \ 54$ \\
                             & $0.40\pm 0.36\ (0.31)$ & $-0.648\pm 0.038$ & $-4.134\pm 0.108$ & $6.62\pm 0.24$ &         & $600, \ 54$ \\
                             & $0.35\pm 0.23\ (0.34)$ & $-0.887\pm 0.037$ & $-4.741\pm 0.131$ & $7.39\pm 0.25$ &         & $1200,\ 54$ \\
        \hline
        AR13470 - Region A   & $0.39\pm 0.39\ (0.24)$ & $-0.362\pm 0.093$ & $-3.358\pm 0.197$ & $7.17\pm 0.63$ & Sub-FOV & $300, \ 27$ \\
        enhanced network     & $0.34\pm 0.29\ (0.21)$ & $-0.316\pm 0.082$ & $-3.628\pm 0.118$ & $5.83\pm 0.33$ &         & $600, \ 27$ \\
                             & $0.28\pm 0.19\ (0.18)$ & $-0.242\pm 0.084$ & $-3.494\pm 0.081$ & $4.74\pm 0.24$ &         & $1200,\ 27$ \\
                             & $0.38\pm 0.36\ (0.24)$ & $-0.355\pm 0.077$ & $-4.000\pm 0.123$ & $6.12\pm 0.30$ &         & $300, \ 54$ \\
                             & $0.33\pm 0.26\ (0.21)$ & $-0.308\pm 0.063$ & $-4.480\pm 0.077$ & $5.36\pm 0.16$ &         & $600, \ 54$ \\
                             & $0.28\pm 0.17\ (0.18)$ & $-0.259\pm 0.069$ & $-4.397\pm 0.062$ & $4.63\pm 0.13$ &         & $1200,\ 54$ \\
        \\
                             & $0.25\pm 0.20\ (0.12)$ & $-0.318\pm 0.113$ & $-3.123\pm 0.159$ & $5.78\pm 0.58$ & enhanced network & $300, \ 27$ \\
                             & $0.22\pm 0.15\ (0.12)$ & $-0.270\pm 0.121$ & $-3.321\pm 0.124$ & $4.89\pm 0.40$ & $B>100\ \rm{G}$  & $600, \ 27$ \\
                             & $0.19\pm 0.12\ (0.12)$ & $-0.212\pm 0.209$ & $-3.197\pm 0.154$ & $4.07\pm 0.51$ &                  & $1200,\ 27$ \\
                             & $0.24\pm 0.15\ (0.12)$ & $-0.308\pm 0.082$ & $-4.049\pm 0.100$ & $5.34\pm 0.24$ &                  & $300, \ 54$ \\
                             & $0.22\pm 0.13\ (0.12)$ & $-0.313\pm 0.096$ & $-4.388\pm 0.095$ & $4.84\pm 0.20$ &                  & $600, \ 54$ \\
                             & $0.19\pm 0.12\ (0.12)$ & $-0.293\pm 0.152$ & $-4.195\pm 0.118$ & $4.28\pm 0.27$ &                  & $1200,\ 54$ \\
        \hline
        AR13470 - Region B   & $0.29\pm 0.37\ (0.15)$ & $-0.322\pm 0.081$ & $-3.102\pm 0.140$ & $6.47\pm 0.51$ & Sub-FOV & $300, \ 27$ \\
        plage                & $0.24\pm 0.25\ (0.09)$ & $-0.273\pm 0.082$ & $-3.308\pm 0.103$ & $5.44\pm 0.33$ &         & $600, \ 27$ \\
                             & $0.17\pm 0.14\ (0.06)$ & $-0.208\pm 0.151$ & $-3.132\pm 0.121$ & $4.25\pm 0.41$ &         & $1200,\ 27$ \\
                             & $0.28\pm 0.31\ (0.12)$ & $-0.315\pm 0.055$ & $-3.907\pm 0.075$ & $5.65\pm 0.19$ &         & $300, \ 54$ \\
                             & $0.23\pm 0.21\ (0.09)$ & $-0.266\pm 0.052$ & $-4.294\pm 0.059$ & $5.13\pm 0.13$ &         & $600, \ 54$ \\
                             & $0.17\pm 0.13\ (0.06)$ & $-0.273\pm 0.086$ & $-4.304\pm 0.074$ & $4.53\pm 0.16$ &         & $1200,\ 54$ \\
        \\
                             & $0.17\pm 0.14\ (0.12)$ & $-0.276\pm 0.087$ & $-2.863\pm 0.106$ & $5.35\pm 0.44$ & plage            & $300, \ 27$ \\
                             & $0.14\pm 0.11\ (0.09)$ & $-0.223\pm 0.141$ & $-2.881\pm 0.131$ & $4.59\pm 0.51$ & $B<-200\ \rm{G}$ & $600, \ 27$ \\
                             & $0.10\pm 0.06\ (0.06)$ & $-0.165\pm 0.328$ & $-2.674\pm 0.205$ & $3.64\pm 0.85$ &                  & $1200,\ 27$ \\
                             & $0.17\pm 0.12\ (0.12)$ & $-0.278\pm 0.064$ & $-3.900\pm 0.069$ & $5.01\pm 0.17$ &                  & $300, \ 54$ \\
                             & $0.14\pm 0.10\ (0.09)$ & $-0.286\pm 0.081$ & $-4.055\pm 0.076$ & $4.70\pm 0.18$ &                  & $600, \ 54$ \\
                             & $0.10\pm 0.06\ (0.06)$ & $-0.152\pm 0.190$ & $-3.968\pm 0.135$ & $4.08\pm 0.32$ &                  & $1200,\ 54$ \\
        \hline
        AR13470 - Region C   & $0.25\pm 0.37\ (0.09)$ & $-0.305\pm 0.083$ & $-2.886\pm 0.116$ & $5.77\pm 0.48$ & Sub-FOV & $300, \ 27$ \\
        plage w/ pores       & $0.21\pm 0.26\ (0.09)$ & $-0.255\pm 0.104$ & $-3.014\pm 0.113$ & $5.00\pm 0.42$ &         & $600, \ 27$ \\
                             & $0.15\pm 0.15\ (0.06)$ & $-0.183\pm 0.193$ & $-2.859\pm 0.153$ & $4.19\pm 0.59$ &         & $1200,\ 27$ \\
                             & $0.24\pm 0.31\ (0.09)$ & $-0.298\pm 0.072$ & $-3.694\pm 0.081$ & $5.14\pm 0.22$ &         & $300, \ 54$ \\
                             & $0.20\pm 0.22\ (0.09)$ & $-0.247\pm 0.070$ & $-4.017\pm 0.067$ & $4.75\pm 0.16$ &         & $600, \ 54$ \\
                             & $0.15\pm 0.13\ (0.06)$ & $-0.173\pm 0.089$ & $-4.074\pm 0.073$ & $4.39\pm 0.17$ &         & $1200,\ 54$ \\
        \\
                             & $0.14\pm 0.16\ (0.09)$ & $-0.267\pm 0.083$ & $-2.782\pm 0.104$ & $5.46\pm 0.45$ & plage            & $300, \ 27$ \\
                             & $0.12\pm 0.09\ (0.06)$ & $-0.212\pm 0.126$ & $-2.784\pm 0.128$ & $4.82\pm 0.53$ & $B<-200\ \rm{G}$ & $600, \ 27$ \\
                             & $0.09\pm 0.06\ (0.03)$ & $-0.141\pm 0.293$ & $-2.600\pm 0.217$ & $4.00\pm 0.94$ &                  & $1200,\ 27$ \\
                             & $0.14\pm 0.12\ (0.09)$ & $-0.260\pm 0.081$ & $-3.601\pm 0.081$ & $4.81\pm 0.23$ &                  & $300, \ 54$ \\
                             & $0.12\pm 0.09\ (0.06)$ & $-0.223\pm 0.069$ & $-3.958\pm 0.065$ & $4.68\pm 0.16$ &                  & $600, \ 54$ \\
                             & $0.09\pm 0.06\ (0.03)$ & $-0.128\pm 0.167$ & $-3.972\pm 0.136$ & $4.38\pm 0.32$ &                  & $1200,\ 54$ \\
        \\
                             & $0.27\pm 0.22\ (0.15)$ & $-0.310\pm 0.139$ & $-3.303\pm 0.202$ & $5.61\pm 0.65$ & emerging flux    & $300, \ 27$ \\
                             & $0.26\pm 0.20\ (0.15)$ & $-0.256\pm 0.155$ & $-3.580\pm 0.191$ & $5.14\pm 0.53$ & $B>100\ \rm{G}$  & $600, \ 27$ \\
                             & $0.20\pm 0.16\ (0.09)$ & $-0.183\pm 0.218$ & $-3.366\pm 0.195$ & $4.32\pm 0.58$ &                  & $1200,\ 27$ \\
                             & $0.27\pm 0.22\ (0.15)$ & $-0.305\pm 0.123$ & $-3.976\pm 0.144$ & $5.03\pm 0.35$ &                  & $300, \ 54$ \\
                             & $0.26\pm 0.19\ (0.15)$ & $-0.340\pm 0.147$ & $-4.464\pm 0.168$ & $4.98\pm 0.34$ &                  & $600, \ 54$ \\
                             & $0.20\pm 0.16\ (0.09)$ & $-0.455\pm 0.160$ & $-4.761\pm 0.181$ & $4.96\pm 0.35$ &                  & $1200,\ 54$ \\
        \hline
        AR13470 - Region D   & $0.30\pm 0.37\ (0.12)$ & $-0.330\pm 0.083$ & $-3.107\pm 0.146$ & $6.53\pm 0.54$ & Sub-FOV & $300, \ 27$ \\
        plage                & $0.25\pm 0.26\ (0.12)$ & $-0.284\pm 0.091$ & $-3.252\pm 0.107$ & $5.24\pm 0.36$ &         & $600, \ 27$ \\
                             & $0.18\pm 0.14\ (0.09)$ & $-0.216\pm 0.166$ & $-3.057\pm 0.134$ & $4.26\pm 0.47$ &         & $1200,\ 27$ \\
                             & $0.28\pm 0.31\ (0.12)$ & $-0.323\pm 0.065$ & $-3.883\pm 0.090$ & $5.67\pm 0.23$ &         & $300, \ 54$ \\
                             & $0.24\pm 0.22\ (0.12)$ & $-0.331\pm 0.061$ & $-4.317\pm 0.069$ & $5.13\pm 0.15$ &         & $600, \ 54$ \\
                             & $0.18\pm 0.13\ (0.09)$ & $-0.400\pm 0.100$ & $-4.364\pm 0.098$ & $4.81\pm 0.22$ &         & $1200,\ 54$ \\
        \\
                             & $0.16\pm 0.15\ (0.09)$ & $-0.285\pm 0.082$ & $-2.808\pm 0.099$ & $5.32\pm 0.43$ & plage            & $300, \ 27$ \\
                             & $0.14\pm 0.11\ (0.09)$ & $-0.232\pm 0.115$ & $-2.806\pm 0.106$ & $4.55\pm 0.43$ & $B<-200\ \rm{G}$ & $600, \ 27$ \\
                             & $0.11\pm 0.08\ (0.09)$ & $-0.172\pm 0.303$ & $-2.588\pm 0.203$ & $3.76\pm 0.89$ &                  & $1200,\ 27$ \\
                             & $0.15\pm 0.12\ (0.12)$ & $-0.285\pm 0.057$ & $-3.792\pm 0.060$ & $4.99\pm 0.16$ &                  & $300, \ 54$ \\
                             & $0.14\pm 0.10\ (0.09)$ & $-0.280\pm 0.082$ & $-4.111\pm 0.064$ & $4.84\pm 0.15$ &                  & $600, \ 54$ \\
                             & $0.11\pm 0.07\ (0.06)$ & $-0.159\pm 0.161$ & $-3.877\pm 0.120$ & $4.19\pm 0.29$ &                  & $1200,\ 54$ \\
                             
	\end{tabular}%
    }
\end{table*}

\section{Coronal oscillation analyses: full report}
\label{sec:appendix_coronal_results_all}
For a better readability we gather here some of the coronal oscillation analyses discussed in the core text.

\begin{figure*}
    \centering
    \includegraphics[width=1.6\columnwidth]{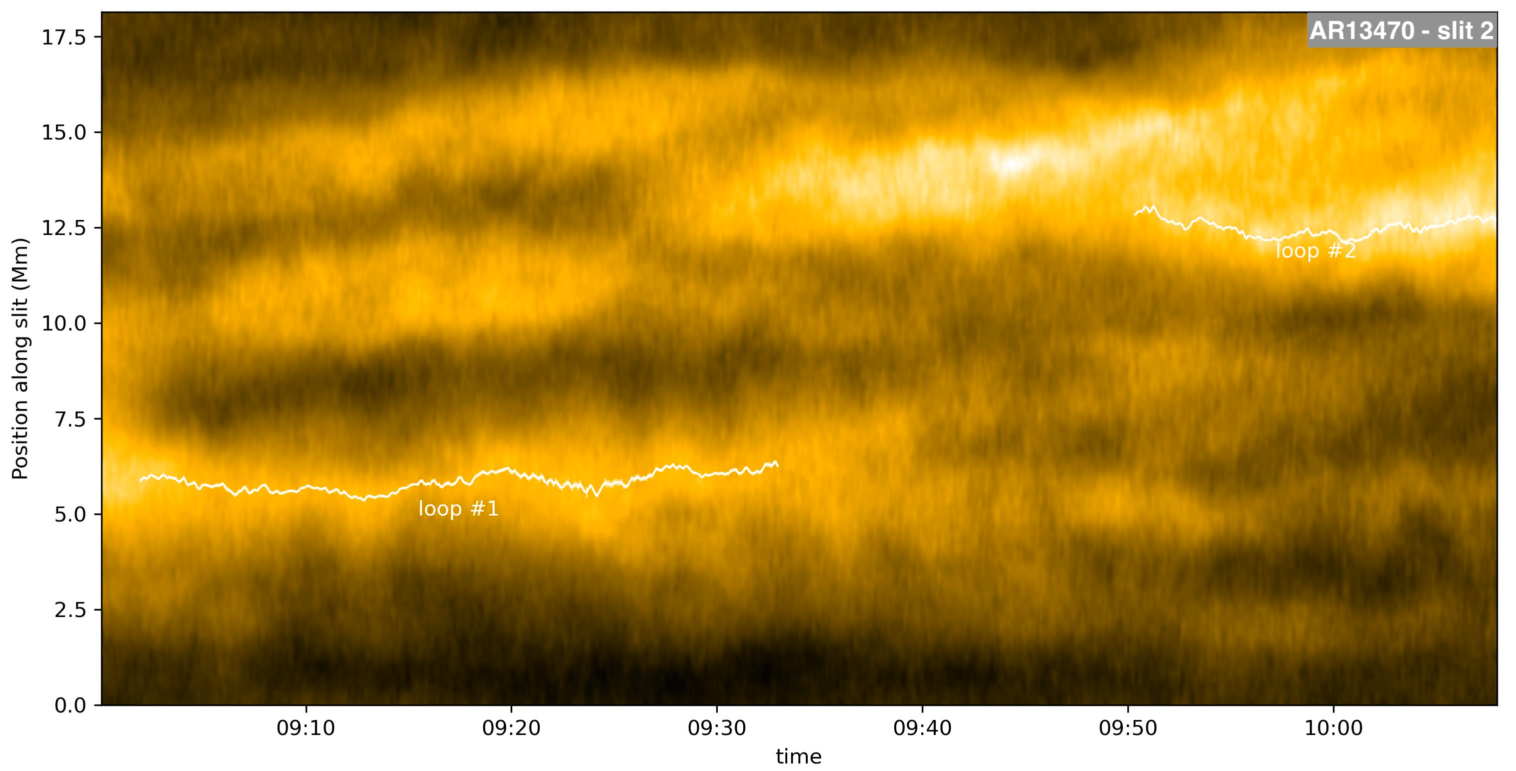}
	\includegraphics[width=2.\columnwidth]{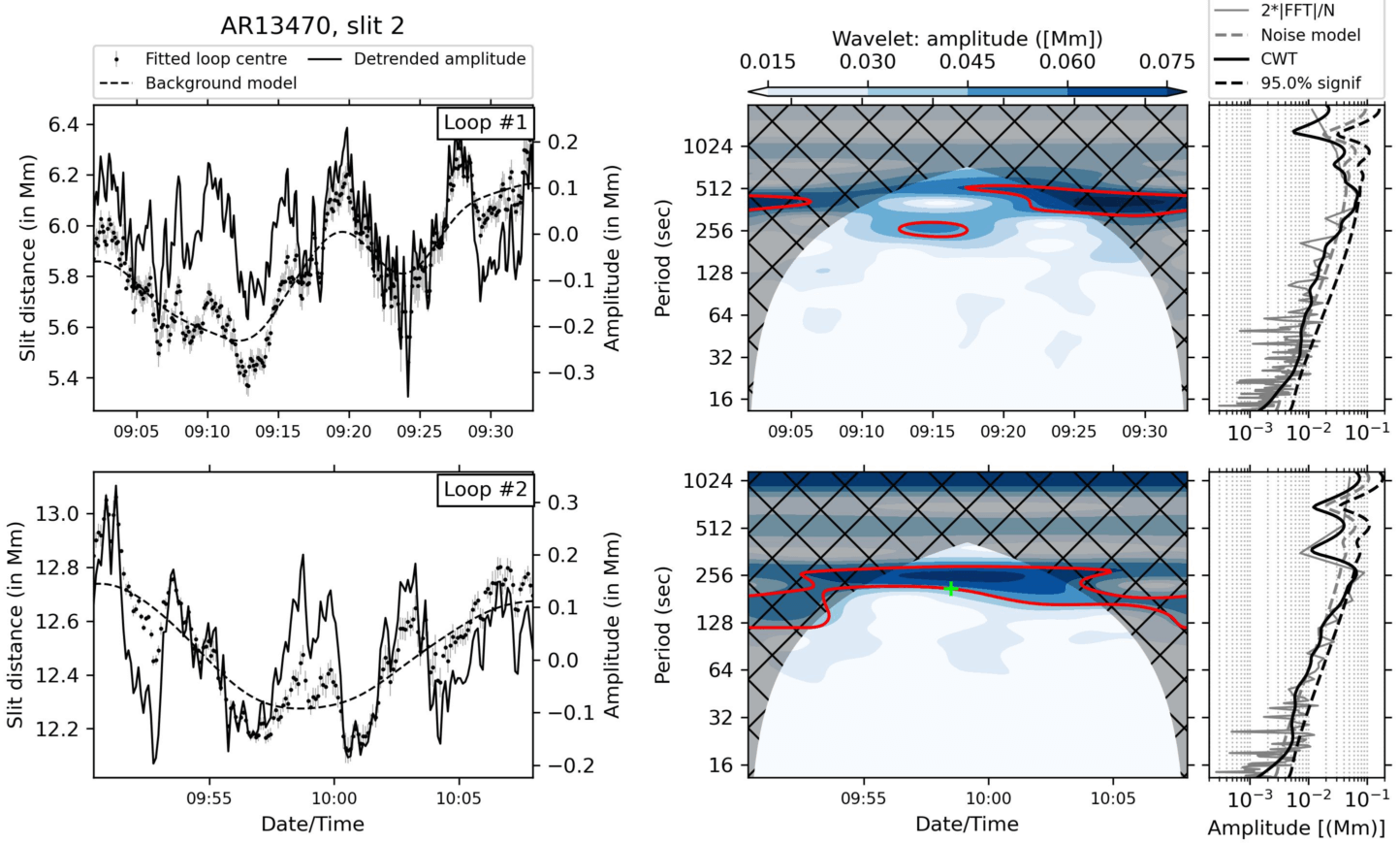}
    \caption{Coronal oscillation analysis for the plage-plage coronal loops of AR13470; slit 2. Top panel: time-distance map of the EUI/HRI intensity along slit 2 along with the fitted loops (solid white lines). Lower panels: the corresponding wavelet analyses following the same format as in Fig.~\ref{fig:2023-10-19_plage-enhanced_allwavelets_slit1}.}
	\label{fig:2023-10-19_plage-plage_allwavelets_slit2a}
\end{figure*}

\begin{figure*}
\centering
\includegraphics[width=1.6\columnwidth]{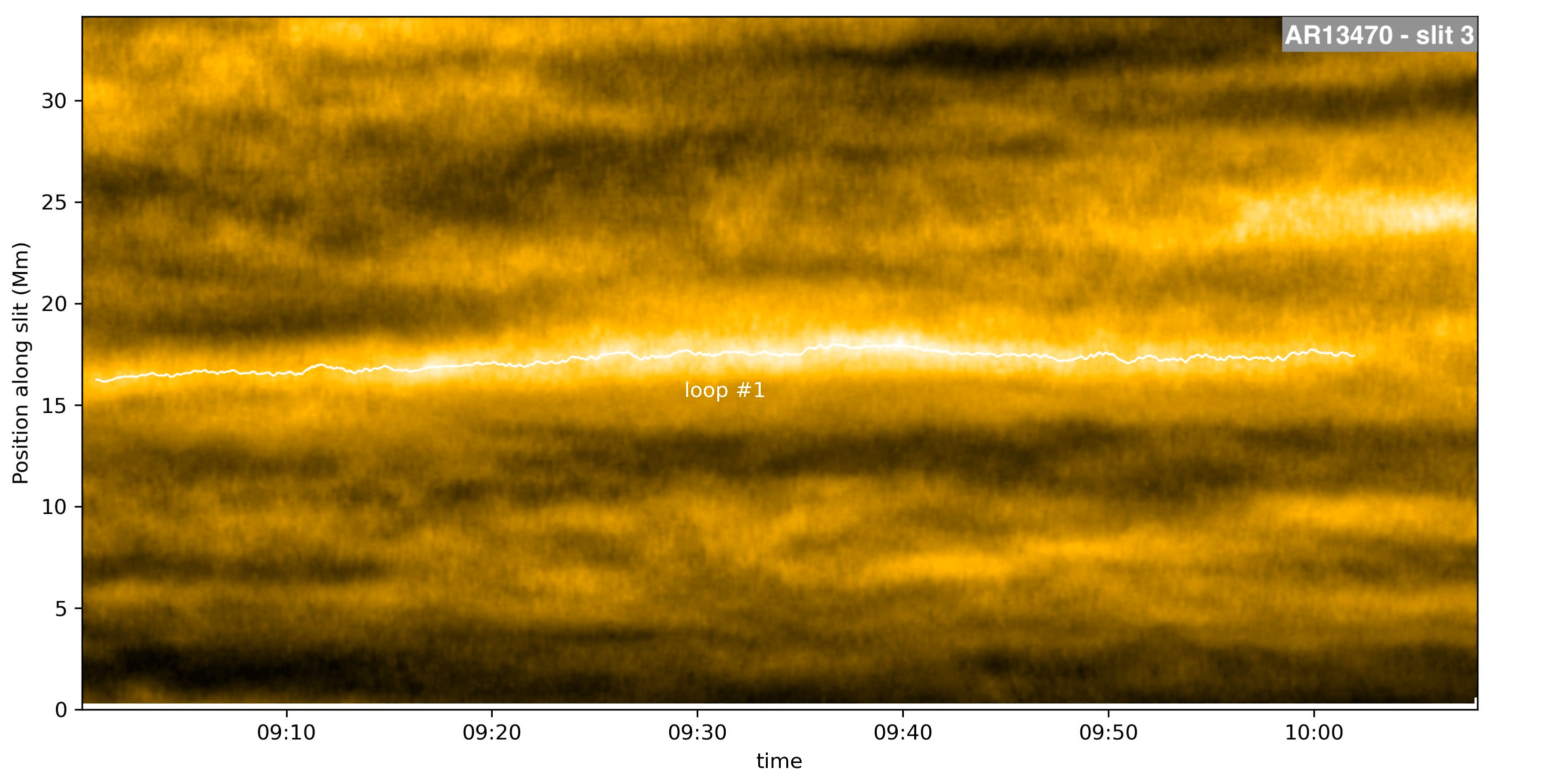}
\includegraphics[width=2.\columnwidth]{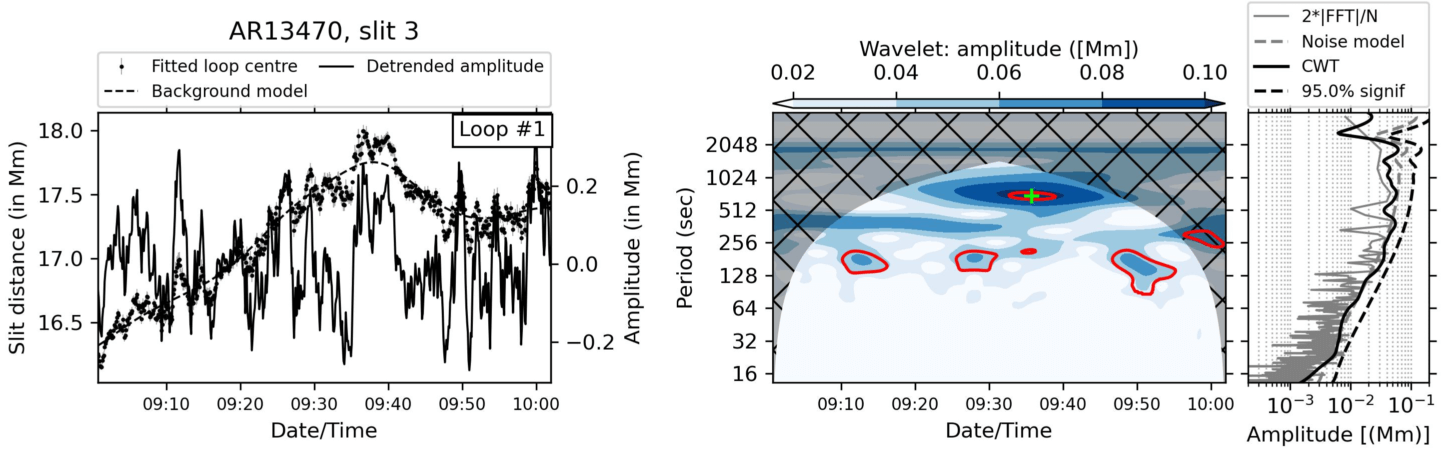}
\caption{Coronal oscillation analysis for the pore-pore coronal loops of AR13470; slit 3. Top panel: time-distance map of the EUI/HRI intensity along slit 3 along with the fitted loops (solid white lines). Lower panels: the corresponding wavelet analyses following the same format as in Fig.~\ref{fig:2023-10-19_plage-enhanced_allwavelets_slit1}.}
\label{fig:2023-10-19_pore-pore_allwavelets_slit3a}
\end{figure*}

\begin{figure*}
    \centering
    \includegraphics[width=1.5\columnwidth]{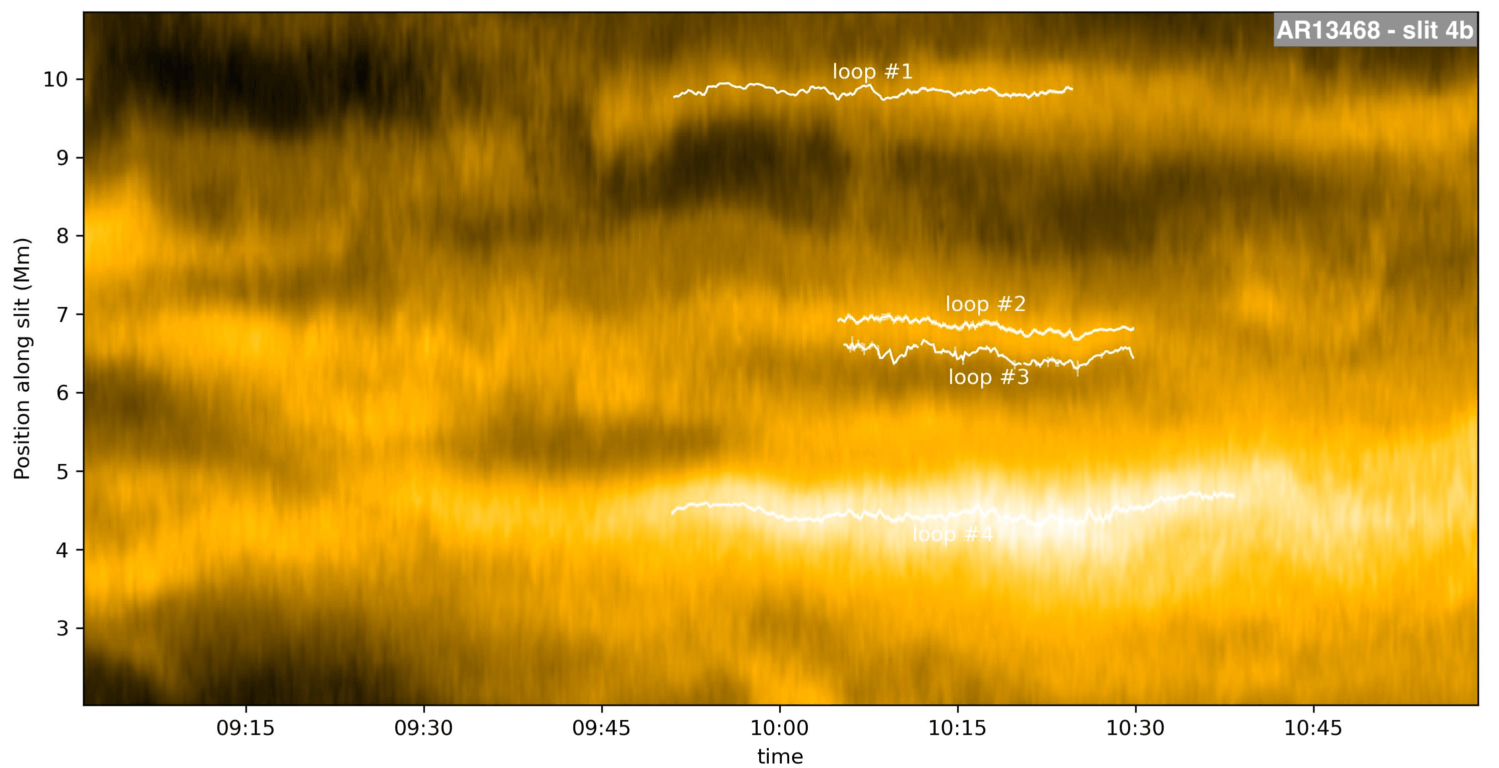}
	\includegraphics[width=1.6\columnwidth]{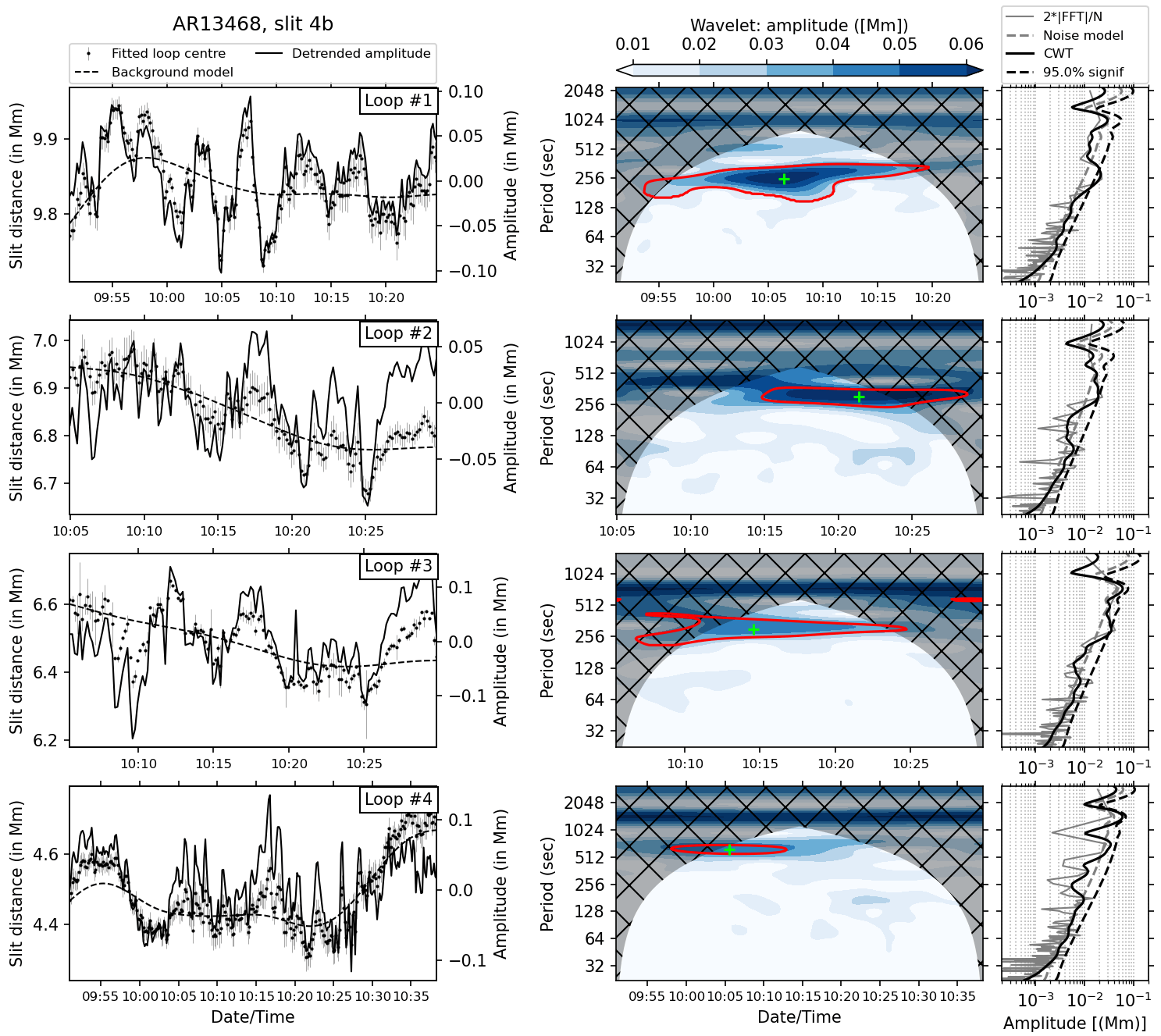}
	\caption{Coronal oscillation analyses for the loops connecting to the sunspot of AR13468; slit 4b. Top panel: time-distance map of the EUI/HRI intensity along slit 4b along with the fitted loops (solid white lines). Lower panels: the corresponding wavelet analyses following the same format as in Fig.~\ref{fig:2023-10-19_plage-enhanced_allwavelets_slit1}.}
	\label{fig:2023-10-20_sunspot_allwavelets_slit2}
\end{figure*}

\begin{figure*}
    \centering
    \includegraphics[width=1.5\columnwidth]{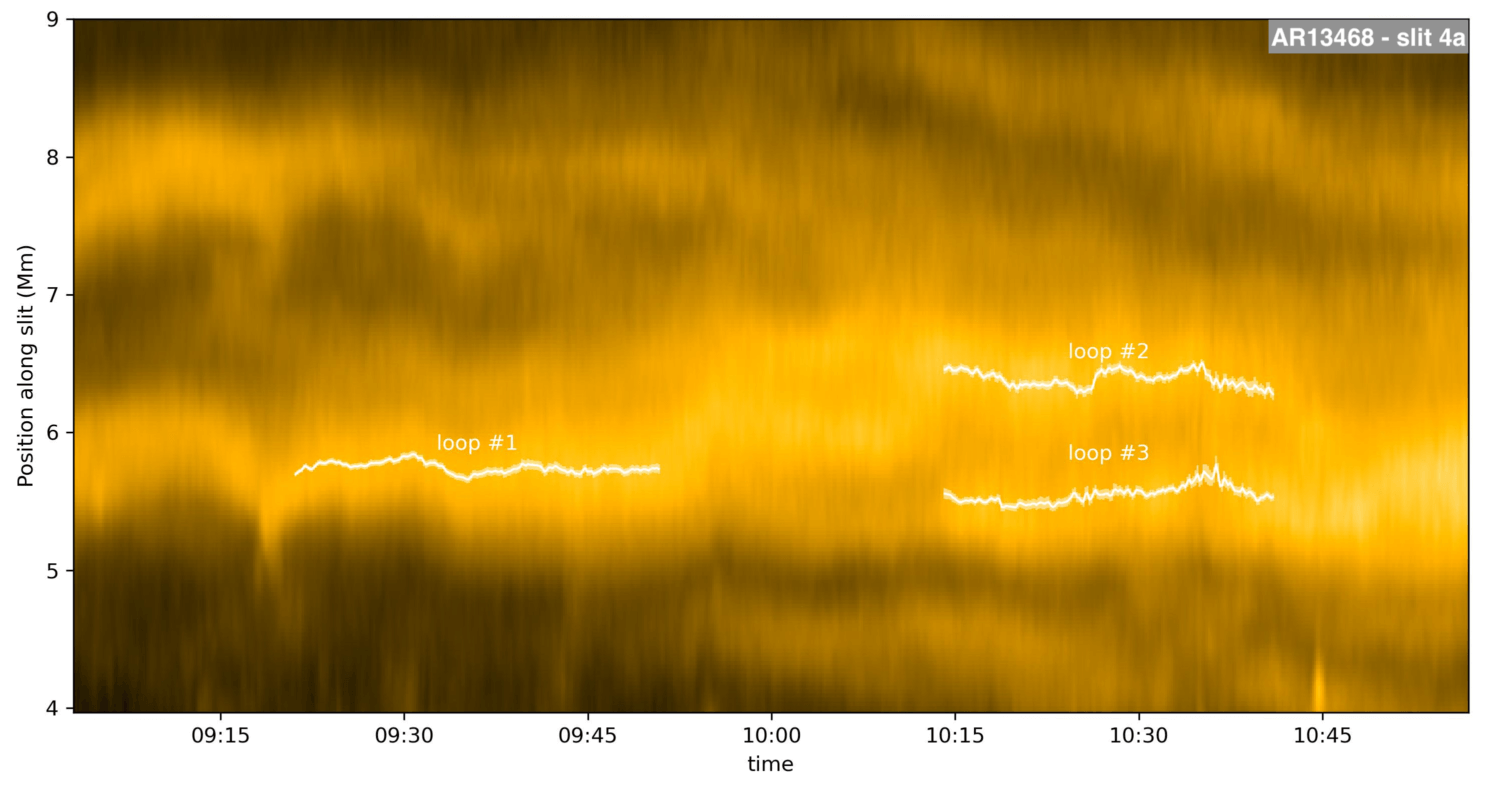}
	\includegraphics[width=2.\columnwidth]{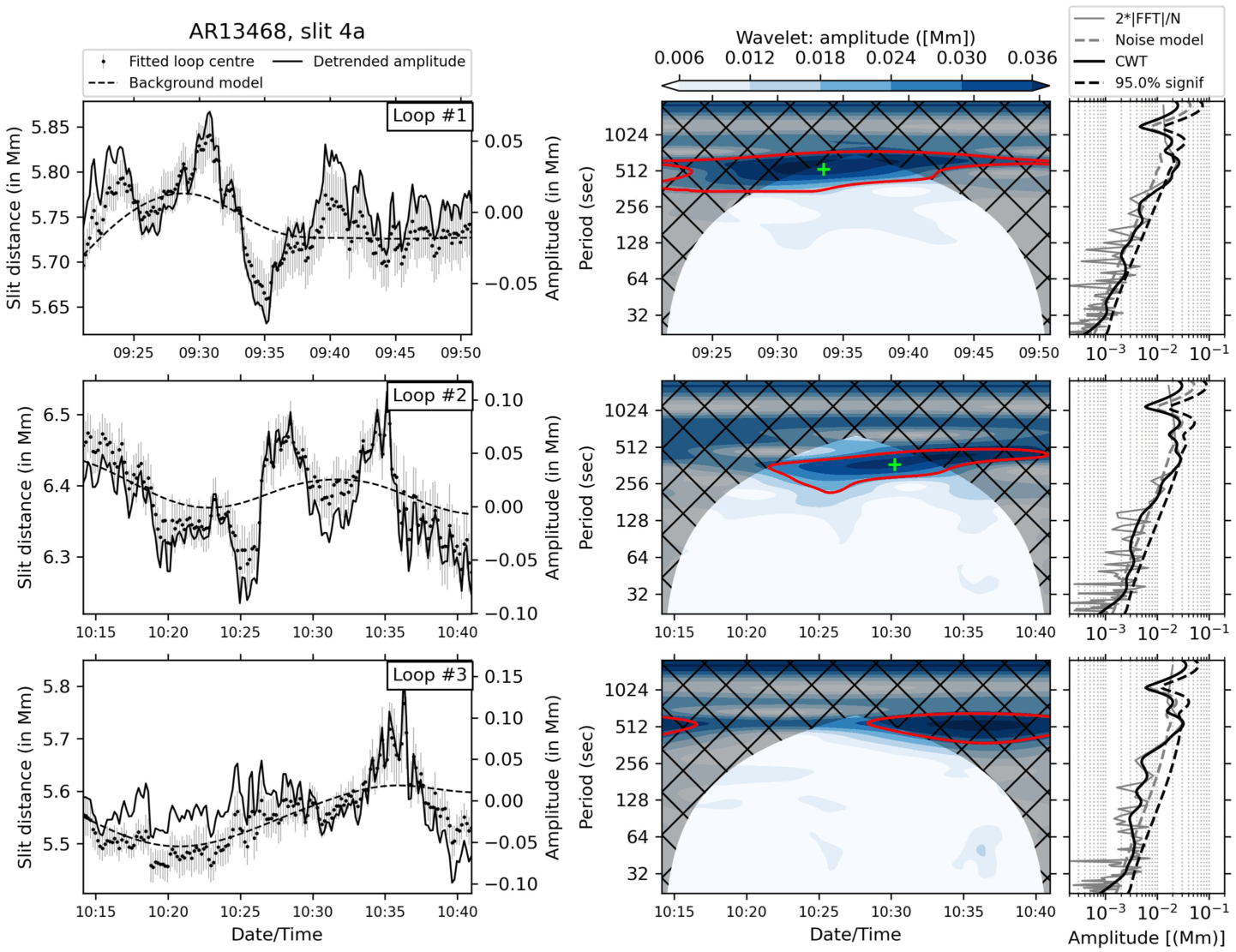}
	\caption{Coronal oscillation analyses for the loops connected to the sunspot of AR13468; slit 4a. Top panel: time-distance map of the EUI/HRI intensity along slit 4a along with the fitted loops (solid white lines). Lower panels: the corresponding wavelet analyses following the same format as in Fig.~\ref{fig:2023-10-19_plage-enhanced_allwavelets_slit1}.}
	\label{fig:2023-10-20_sunspot_allwavelets_slit1}
\end{figure*}

\begin{figure*}
    \centering
    \includegraphics[width=1.5\columnwidth]{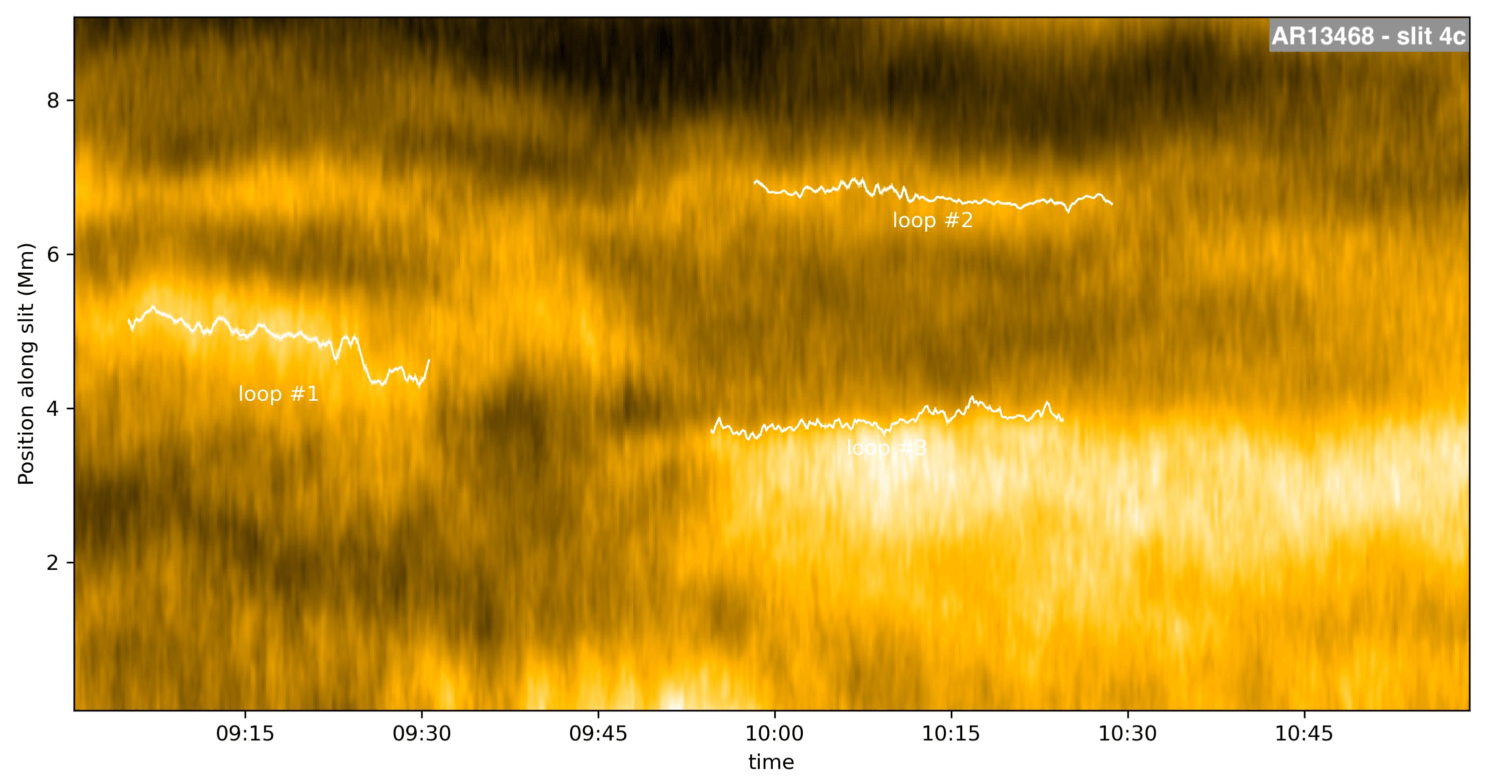}
	\includegraphics[width=2.\columnwidth]{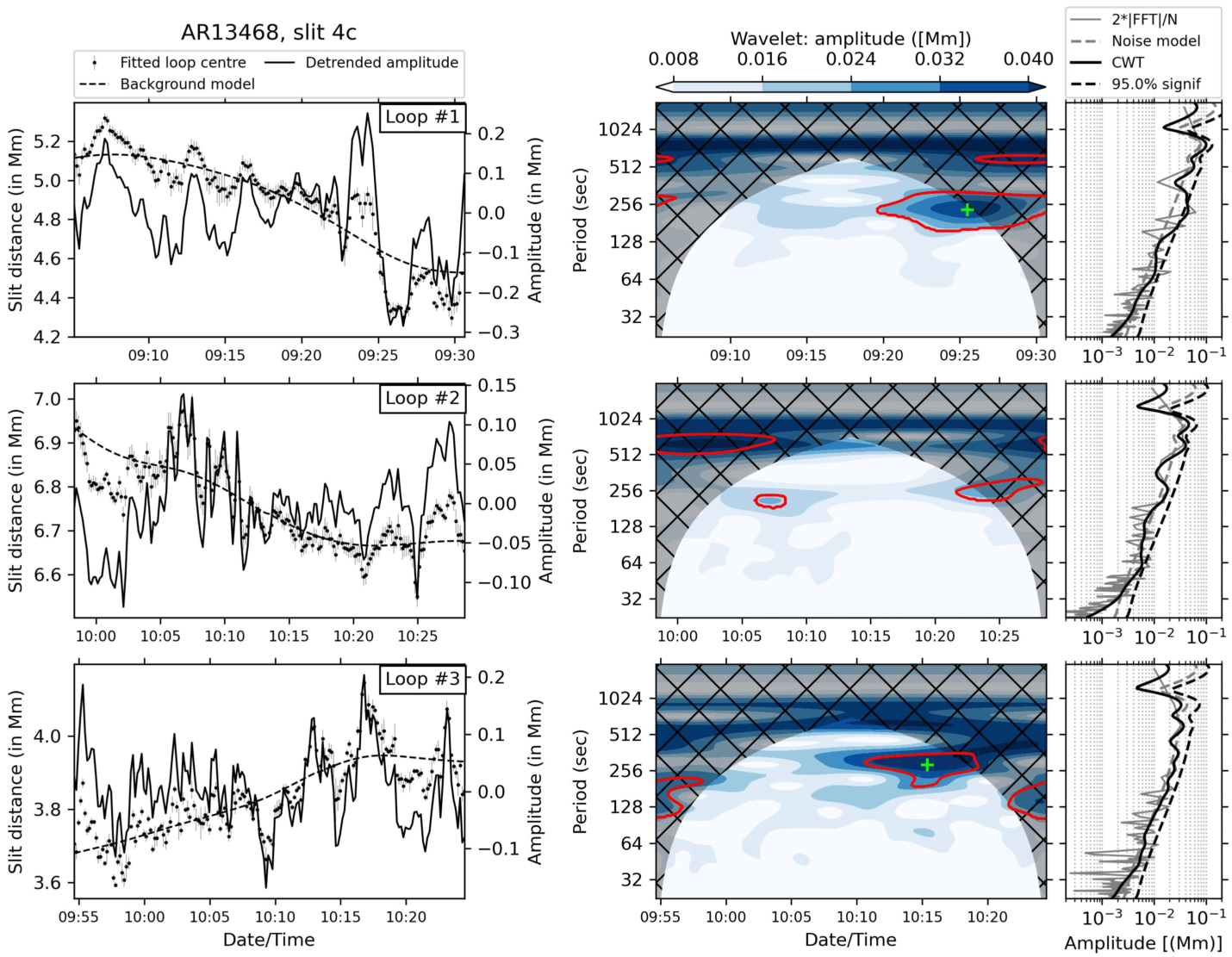}
	\caption{Coronal oscillation analyses for the loops connected to the sunspot of AR13468; slit 4c. Top panel: time-distance map of the EUI/HRI intensity along slit 4c along with the fitted loops (solid white lines). Lower panels: the corresponding wavelet analyses following the same format as in Fig.~\ref{fig:2023-10-19_plage-enhanced_allwavelets_slit1}.}
	\label{fig:2023-10-20_sunspot_allwavelets_slit3}
\end{figure*}

\end{appendix}

\end{document}